\documentclass[12pt,preprint]{aastex}
\usepackage{lscape}
\usepackage{color}

\shorttitle{Near-IR Spectroscopy of Seyferts.}
\shortauthors{Ramos Almeida et al.}

\begin{document}

\title{Near-Infrared Spectroscopy of Seyfert Galaxies. Nuclear Activity and Stellar Population.}

\author{C. Ramos Almeida\altaffilmark{1}, A. M. P\'{e}rez Garc\'\i a\altaffilmark{1}, 
and J. A. Acosta-Pulido\altaffilmark{1}}

\altaffiltext{1}{Instituto de Astrof\'\i sica de Canarias (IAC), 
              C/V\'\i a L\'{a}ctea, s/n, E-38205, La Laguna, Tenerife, Spain.
cra@iac.es, apg@iac.es, jap@iac.es}

\begin{abstract}

Near-infrared spectroscopic data for the five Seyfert galaxies with jet-gas interaction Mrk~348, 
Mrk~573, Mrk~1066, NGC~7212, and NGC~7465, taken with the LIRIS near-infrared
camera/spectrometer at the William Herschel Telescope (WHT) are reported. 
The long-slit spectra reveal the characteristic strong 
emission lines of this type of objects. Many forbidden transitions and hydrogen recombination lines are employed here to study
the excitation and ionization mechanisms that are dominating the narrow-line region emission of these
objects, that is affected by the radio-jet interaction. Several absorption features are also detected in the H and K bands 
of these galaxies, 
allowing us to identify the spectral types that are producing them. We find that the continuum can be 
reproduced by a combination of late-type stellar templates plus a Blackbody component associated to host dust, 
mainly contributing to the K band emission.
The detection of the permitted O I and Fe II lines and broad components 
of the hydrogen recombination lines in the spectra of Mrk~573 and NGC~7465 allows the reclassification of 
these two galaxies that are 
not canonical Type-2 Seyferts: Mrk~573 is confirmed to be an obscured Narrow-line Seyfert 1 and
NGC~7465 is revealed for the first time as a Type-1 LINER through its near-infrared spectrum.

\end{abstract}


\section{Introduction}
\label{intro} 
 
Active galactic nuclei (AGN) have been intensively studied in the optical due to the
widely used ground-based visible spectrographs. However, there are very few detailed 
near-infrared spectroscopic studies of AGN, and
particularly, of Seyfert galaxies. This spectral range offers a wide variety of diagnostic tools 
used to characterize relevant phenomena acting in the narrow line region (NLR), partially ionized media or borders of molecular 
clouds of AGNs. According with the Unified Model \citep{Antonucci93,Urry95}, the differences
between the spectra of Type-1 and Type-2 AGN are only due to orientation effects, as a consequence
of the existence of a blocking structure of dusty material. Ways of unveiling the supposedly hidden broad line region (BLR)
of Type-2 AGN are spectropolarimetric observations \citep{Antonucci85,Tran95a,Tran95b} and infrared spectroscopy. 
Particularly,  
near-infrared spectroscopy is well-suited for detecting broad components in the hydrogen recombination lines through a less
extinct line of sight (LOS). 
On the other hand, the non-stellar emission that dominates the ultraviolet (UV) and optical 
continuum in Type-1 Seyferts no longer dominates the near-infrared emission \citep{Kishimoto05}. Therefore, this range offers
an opportunity to study the stellar content of the nuclear region of Seyferts and reveals signatures of recent
or intense star formation.

The gas motions in the NLR are generally dominated by the 
gravitational field of the galaxy, with the line widths matching the bulge gravitational velocities 
\citep{Whittle85,Wilson85,Whittle92,Nelson96}. However, in Seyferts with relatively strong radio emission there is 
evidence for significant jet-related additional acceleration \citep{Ferruit02,Veilleux02}.
In Seyfert galaxies presenting this jet-gas
interaction, shocks can play several roles. They can compress and sweep the gas, ablate the clouds,
destroy them, or even remain separate \citep{Whittle04}. Shocks are usually determinant not only for the morphology, 
but also for the  kinematics and excitation of the NLR. 
This makes the interaction between the radio-emitting flows and the line-emitting gas an interesting area 
of research in the study of AGN. 
For example, Hubble Space Telescope narrow-band imaging of high spatial 
resolution revealed bow-shock-shaped emission-line regions in sources with jetlike radio structures \citep{Pogge96,Axon98}.

The high-excitation gas is usually confined to a biconical
structure centred on the nucleus, usually seen in Seyfert 2 galaxies. 
The observed morphology of these ionization cones is 
generally well aligned with the radio emission \citep{Allen99} and their detection 
is usually taken as an evidence of the existence of an 
inner toroidal distribution of obscuring material \citep{Mulchaey96}. 
The dominating excitation mechanism of the gas in the cones is the nuclear photoionization. However, the interaction 
with fast shocks seems 
to play an important role in the excitation of the NLR of Seyfert galaxies \citep{Dopita95,Dopita96,Rosario04,Ramos06}.
In particular, for the case of Mrk~78, which is a classical example of a jet-gas interating AGN with large bipolar flows
\citep{Whittle88,Pedlar89}, \citet{Ramos06} presented evidences of the radio jet influence based on the detection of  
enhanced [Fe II] emission and double-peaked profiles in the spectrum of the outer regions. In a similar study, 
\citet{Jackson07} reported
the existence of an extranuclear region with radio jet interaction in Mrk~34, where the [Fe II] is considerably enhanced.

We present here near-infrared spectra of five nearby Seyferts presenting jet-gas interaction and 
optical ionization cones. 
Extended Narrow-Line Region (NLR)
emission has been reported and characterized by other authors for some of the objects (e.g., Mrk~573, 
\citealt{Falcke98} and \citealt{Ferruit99} or Mrk348, \citealt{Alonso98}).
The high-quality of our data permits us to assess the presence of extended line-emission up to hundreds of kiloparsecs 
in most of the galaxies considered here. In this paper, we report only about the nuclear properties.
A detailed study of the extended emission will be the subject of a forthcoming paper (C. Ramos Almeida et al. 2009, 
in preparation). 
Numerous emission features, including both low- and high-ionization lines and molecular transitions 
are present in our nuclear spectra.
We study the acting excitation mechanisms, together
with other properties of the galaxies that can be derived from near-infrared diagnostics (e.g., extinction). 
We analyze the continuum shape and dominant stellar populations towards the nucleus.

No previous high-quality near-infrared spectroscopy is available in the
literature for these galaxies, with the exceptions of Mrk~1066, observed by \citet{Knop01} 
and Mrk~573 together with Mrk~1066, studied by
\citet{Riffel06}.
Indeed, the spectra presented here constitute an homogeneus data set, 
taken with the same instrument and covering the same spectral range. Several emission and absorption
features detected in our spectra are not reported in previous works. Many of these lines, together with the 
detection of a broad pedestal in the Pa$\beta$ profile of Mrk~573, have recently led to reclassification 
of this galaxy as an obscured Narrow-line Seyfert 1 \citep{Ramos08}.
Morphological classification of the galaxies, spectroscopic redshifts, distance, physical scale,
galactic extinction amount \citep{Schlegel98}, and hydrogen density column \citep{Guainazzi05,Awaki06} 
are reported in Table \ref{info}.

\begin{table}[ !ht ]
\centering
\begin{tabular}{lcccccc}
\hline
\hline
Galaxy & morphology & z$_{spec}$ & \multicolumn{1}{c}{distance} & \multicolumn{1}{c}{scale} & \multicolumn{1}{c}{E(B-V)} & \multicolumn{1}{c}{N$_{H}$} \\
 & & & (Mpc) & (pc~arcsec$^{-1}$) & (mag) & (cm$^{-2}$) \\
\hline
Mrk~348   & SA(s)0/a	  & 0.015 $^{a}$  & 60  & 291 & 0.067 & 1.3x10$^{23}$ 		\\
Mrk~573   & (R)SAB(rs)0+  & 0.017 $^{b}$  & 69  & 333 & 0.023 & $>$1.6x10$^{24}$  		 \\ 
Mrk~1066  & (R)SB(s)0+    & 0.012 $^{c}$  & 48  & 233 & 0.132 & 9x10$^{23}$  		\\
NGC~7212  & Sab   	  & 0.027 $^{d}$  & 106 & 516 & 0.072 & $>$1.6x10$^{24}$  		 \\
NGC~7465  & (R')SB(s)0    & 0.007 $^{e}$  & 26  & 127 & 0.078 & 4.6x10$^{23}$  		 \\
\hline  																																											   
\end{tabular}  
\caption{\footnotesize{Morphology of our targets taken from the NASA/IPAC Extragalactic Database (NED), spectroscopic
redshift, distance, physical scale, galactic extinction amount, and hydrogen column density. 
References. (a) \citet{Huchra99}; (b) \citet{Ruiz05}; (c) \citet{Bower95}; (d) \citet{Keel96}; (e) \citet{Lu93}.
Notes. The distance to the galaxies was determined using H$_{0}$ = 75 km s$^{-1}$ Mpc$^{-1}$.}} 
\label{info}
\end{table}

\section{Observations and data reduction}

	Near-infrared spectra in the range 0.8-2.4 \micron~were obtained from 2005 July to 2008 August, 
using the near-infrared camera/spectrometer LIRIS \citep{Manchado04,Acosta-Pulido03}, attached to 
the Cassegrain focus of the 4.2 m William Herschel Telescope (WHT). 
LIRIS is equipped with a Rockwell Hawaii 1024 x 1024 HgCdTe array detector. The spatial scale is
0.25\arcsec~pixel$^{-1}$, and the slit width used during the observations was 0.75\arcsec~(allowing a
spectral resolution of $\sim$450 and 500 km~s$^{-1}$ in the ZJ and HK ranges, respectively) except for the
case of NGC~7212, for which the 1\arcsec~slit ($\sim$600 and 650 km~s$^{-1}$ resolutions in the ZJ and HK ranges,
respectively) was chosen because of the worse value of the seeing. The journal of the observations is reported 
in Table \ref{log}.

\begin{table}[ !ht ]
\centering
\footnotesize
\begin{tabular}{llccccccc}
\hline
\hline
Galaxy & Obs. Date &\multicolumn{2}{c}{Exposure Time}& P.A. &\multicolumn{2}{c}{Airmass} & Seeing & Telluric Star  \\
& & ZJ & HK & & ZJ & HK & &   \\
\hline
Mrk~348 & 2005 July 7     & 12x300 s & 12x250 s & 155$^{o}$  & 1.30 & 1.10  & $\sim$0.8\arcsec     & Hip5671           \\
Mrk~573 & 2006 July 17-18 & 8x500 s  & 8x400 s  & 122$^{o}$  & 1.33 & 1.38  & $\sim$0.8\arcsec     & Hip2047 	    \\
Mrk~1066& 2006 July 18 *  & 3x500 s  & 8x400 s  & 135$^{o}$  & 1.21 & 1.08  & $\sim$0.8/0.7\arcsec & Hip8581 	    \\
NGC~7212& 2006 July 16    & 6x500 s  & 10x400 s & 145$^{o}$  & 1.14 & 1.06  & $\sim$1\arcsec       & Hip113977         \\
NGC~7465& 2005 July 5-6   & 8x300 s  & 12x300 s & 327$^{o}$  & 1.03 & 1.03  & $\sim$0.7\arcsec     & Hip1264/Hip198    \\
\hline	   					 											        																																													  
\end{tabular}	
\caption{\footnotesize{Log of observations. *The HK spectra of Mrk~1066 was taken in 2008 August 12.}}
\label{log} 
\end{table}

Observations were performed following an ABBA telescope-nodding pattern, placing the source in two
positions along the slit, separated by 15\arcsec. Individual frames were taken with the integration times
reported in Table \ref{log} in each of the ZJ and HK ranges. The wavelength calibration was provided by
observation of the argon lamp available in the calibration unit at the A\&G box of the telescope. In order to
obtain the telluric correction and the flux calibration for each galaxy, nearby A0 stars were observed with 
the same configuration and the most similar airmass to the galaxy as possible (see Table \ref{log}).

The data were reduced following stardard procedures for near-infrared spectroscopy, using the 
{\it lirisdr} dedicated software within the IRAF\footnote{IRAF
is distributed by the National Optical Astronomy Observatory, which is operated by the Association of
Universities for the Research in Astronomy, Inc., under cooperative agreement with the National Science
Foundation (http://iraf.noao.edu/).} enviroment. For a detailed description of the reduction process, see
\citet{Ramos06}. Consecutive pairs of AB
two-dimensional spectra were subtracted to remove the sky background. The resulting frames were then
wavelength-calibrated and flat-fielded before registering and co-adding all frames to provide the final
spectra.

In order to study the nuclear emission of the five galaxies here considered, we have extracted
the spectra covering 1.5\arcsec~centered on the maximum of the galaxy profiles. The aperture was
selected attending to the spatial resolution given by the seeing conditions.

\begin{figure}[!htp]
\centering
{\par
\includegraphics[width=8cm,angle=90]{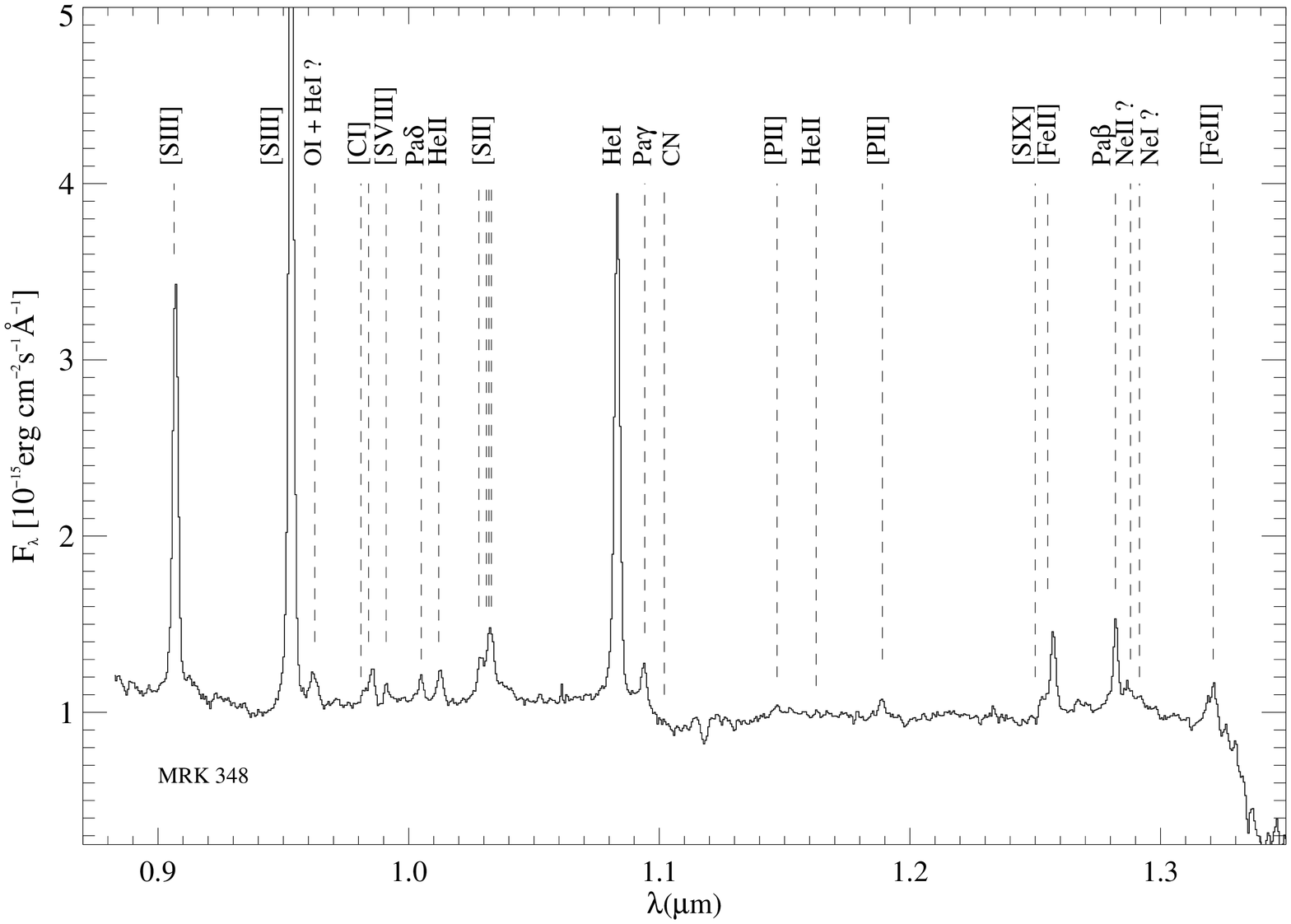}
\includegraphics[width=8cm,angle=90]{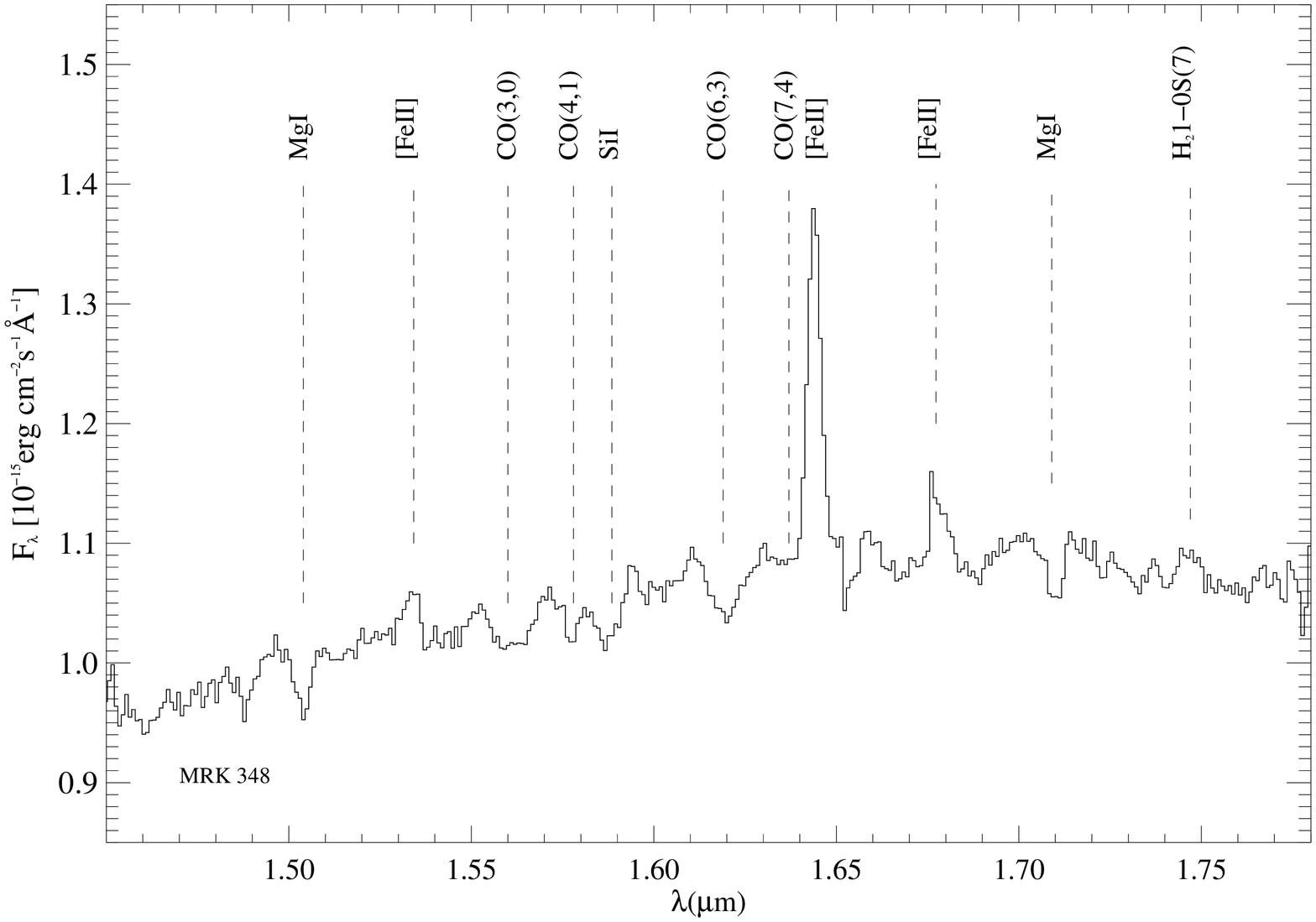}
\includegraphics[width=8cm,angle=90]{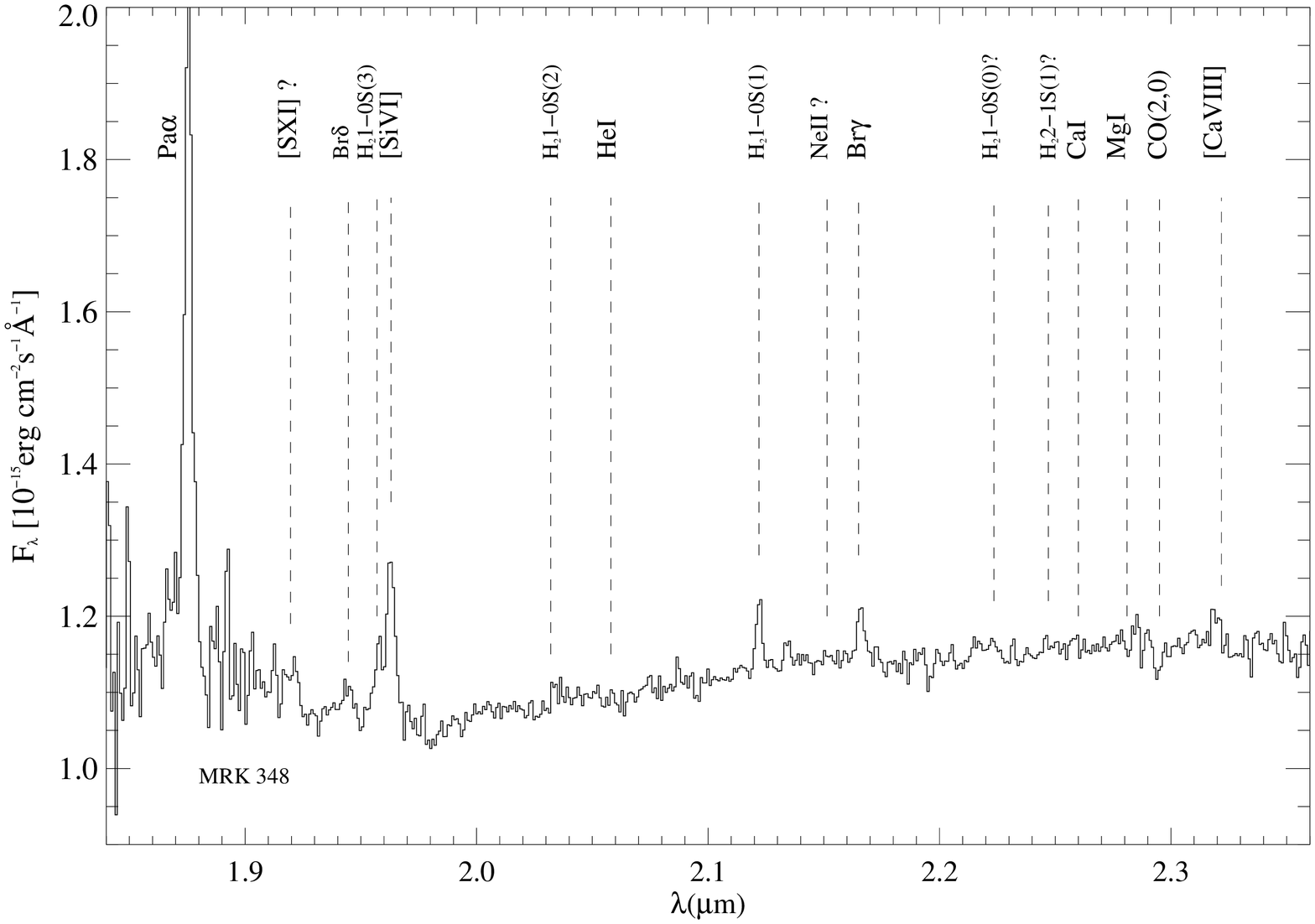}\par}
\figcaption{\footnotesize{Flux-calibrated spectra corresponding to the nuclear region of the galaxy
Mrk~348 within an aperture of 1.5\arcsec, in the ZJ, H, and K ranges. Note the red excess towards long wavelengths.}
\label{fi:m348}}
\end{figure}

\begin{figure}[!htp]
\centering
{\par
\includegraphics[width=8cm,angle=90]{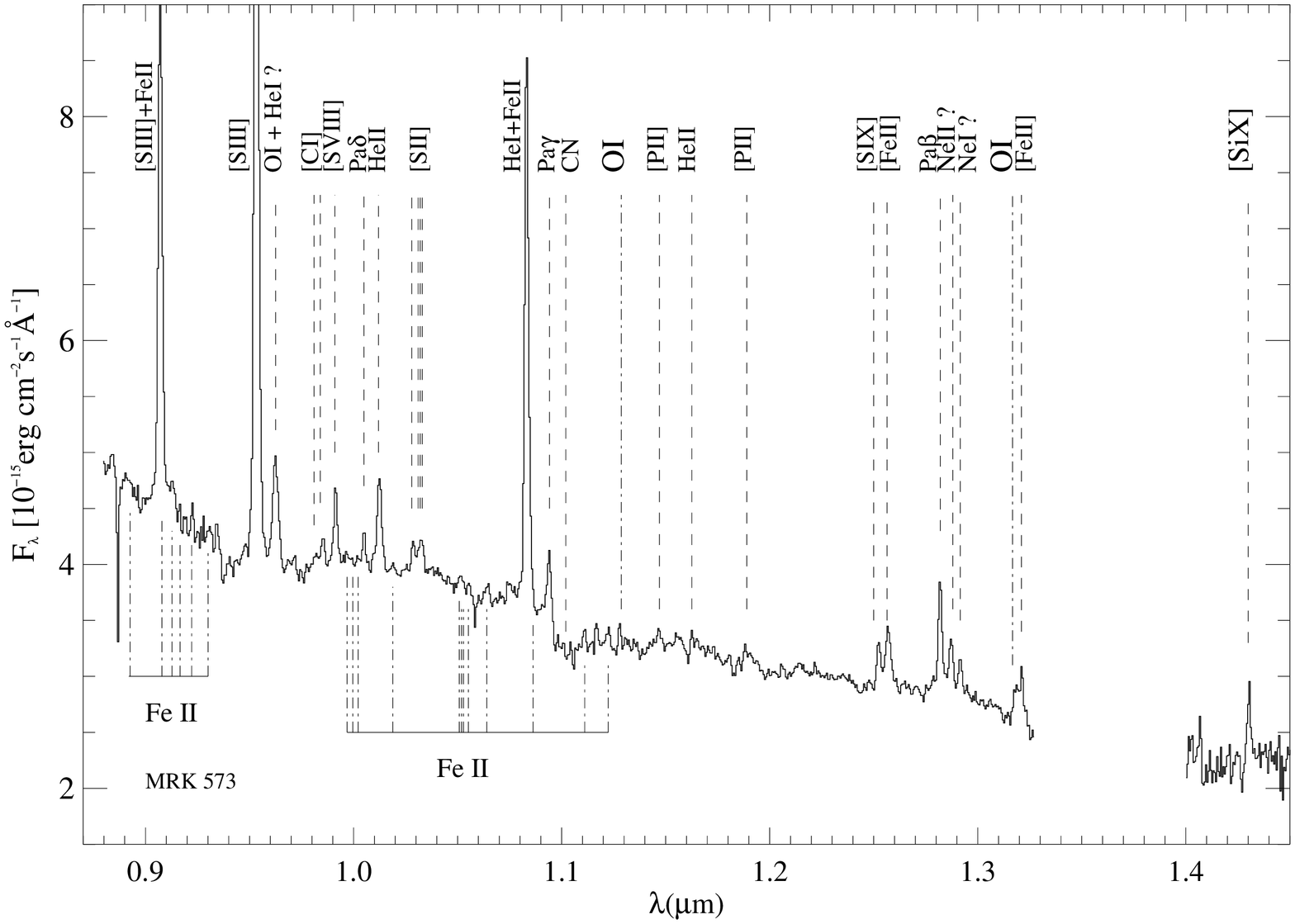}
\includegraphics[width=8cm,angle=90]{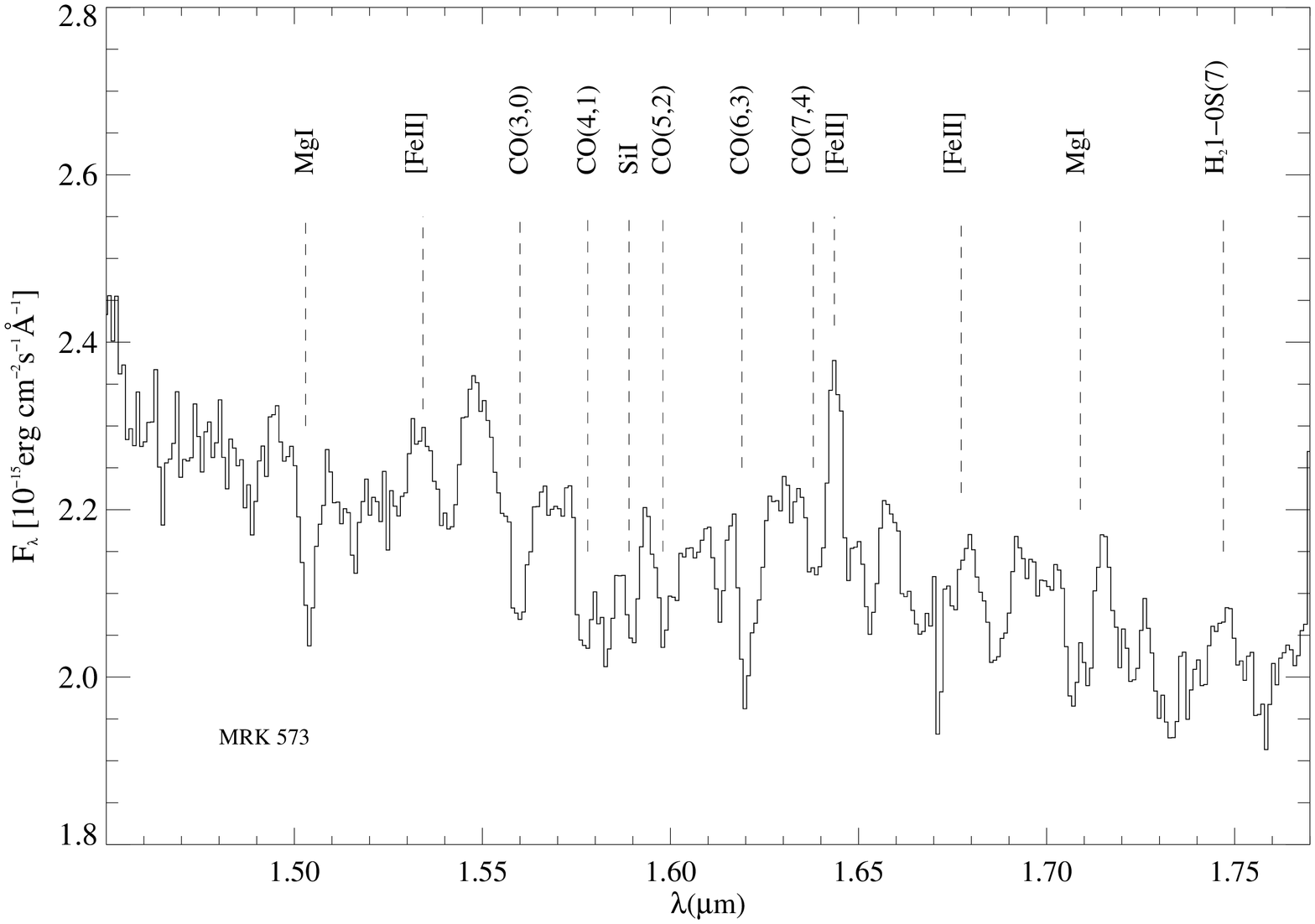}
\includegraphics[width=8cm,angle=90]{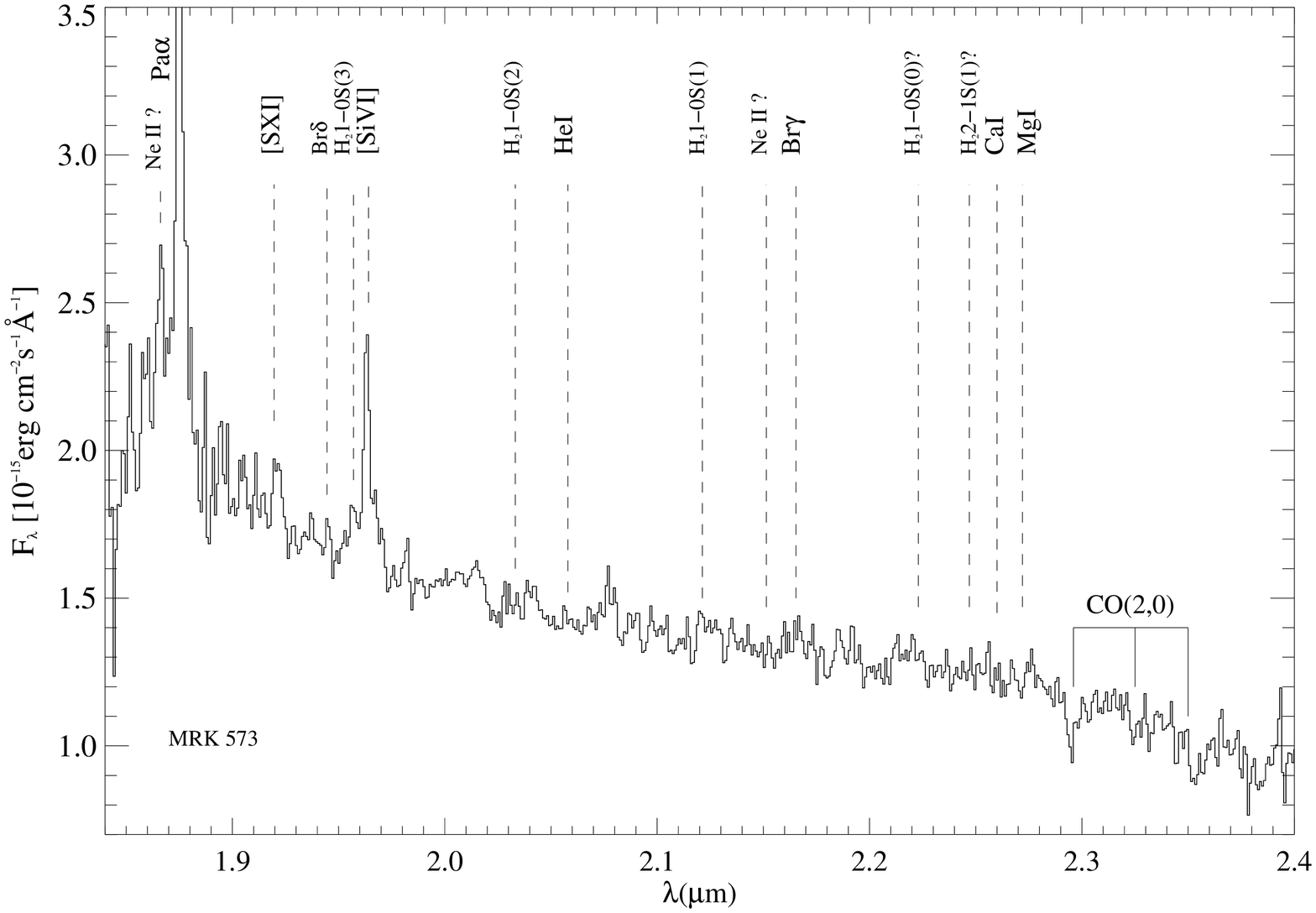}\par}
\figcaption{\footnotesize{Same as in Figure \ref{fi:m348}, but for Mrk~573. Note the broad wings of Pa$\beta$ and 
Pa$\alpha$ as well as the presence of the permitted Fe II and O I lines.}
\label{fi:m573}}
\end{figure}

\begin{figure}[!htp]
\centering
{\par
\includegraphics[width=8cm,angle=90]{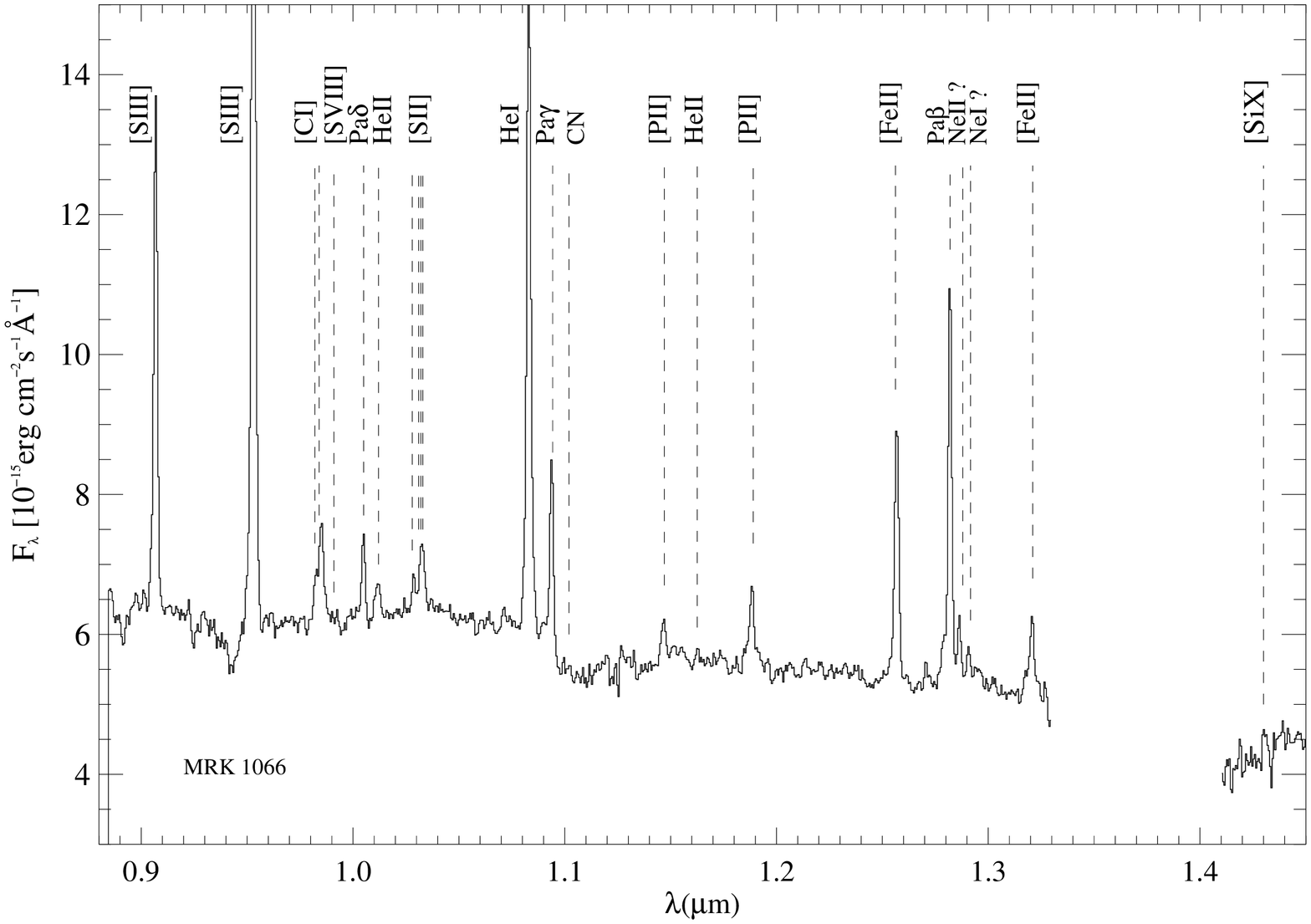}
\includegraphics[width=8cm,angle=90]{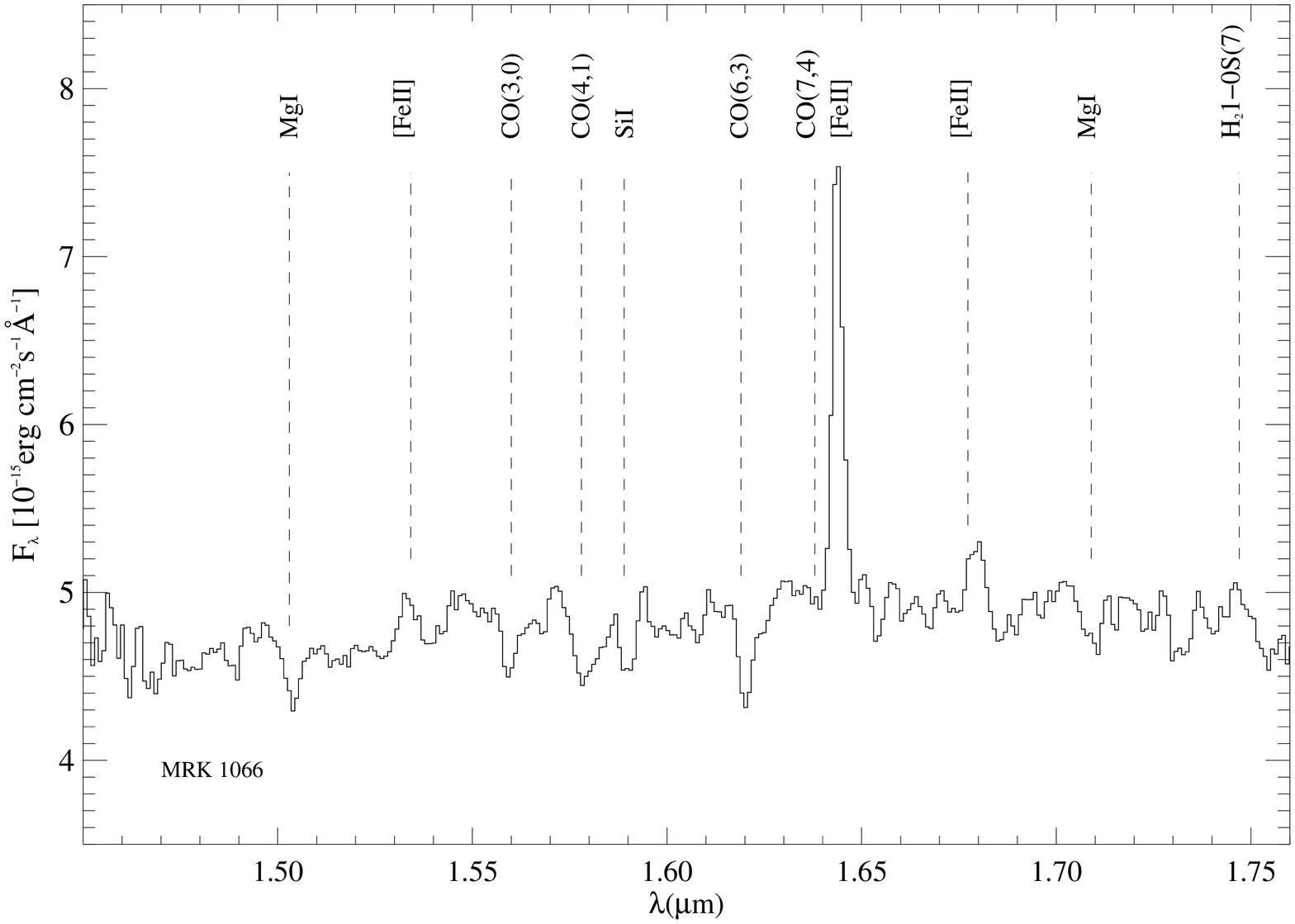}
\includegraphics[width=8cm,angle=90]{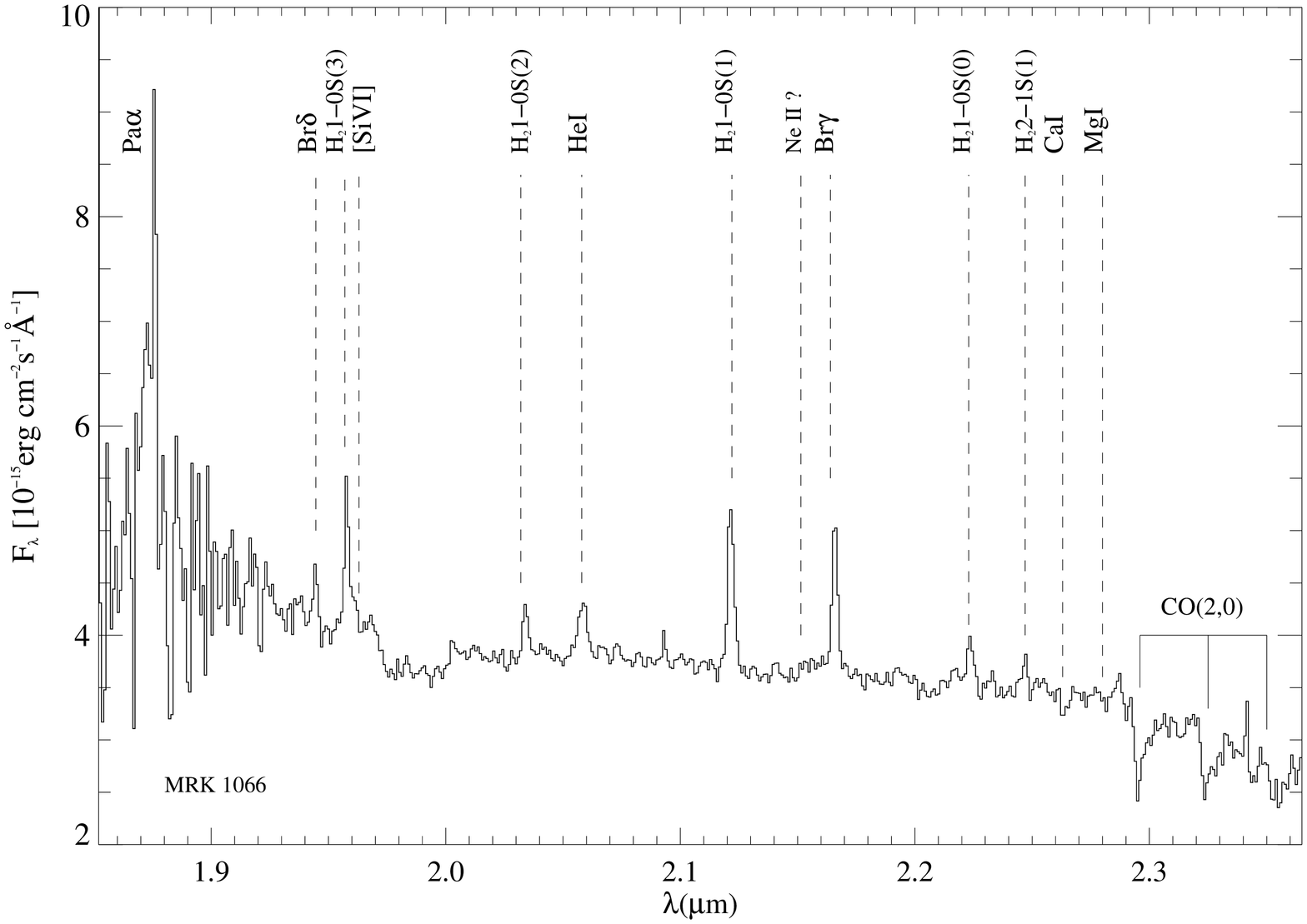}\par}
\figcaption{\footnotesize{Same as in Figure \ref{fi:m348}, but for Mrk~1066. Note the presence of the intense H$_{2}$ lines as
well as the prominent He I$\lambda$2.06 \micron~line.}
\label{fi:m1066}}
\end{figure}

\begin{figure}[!htp]
\centering
{\par
\includegraphics[width=8cm,angle=90]{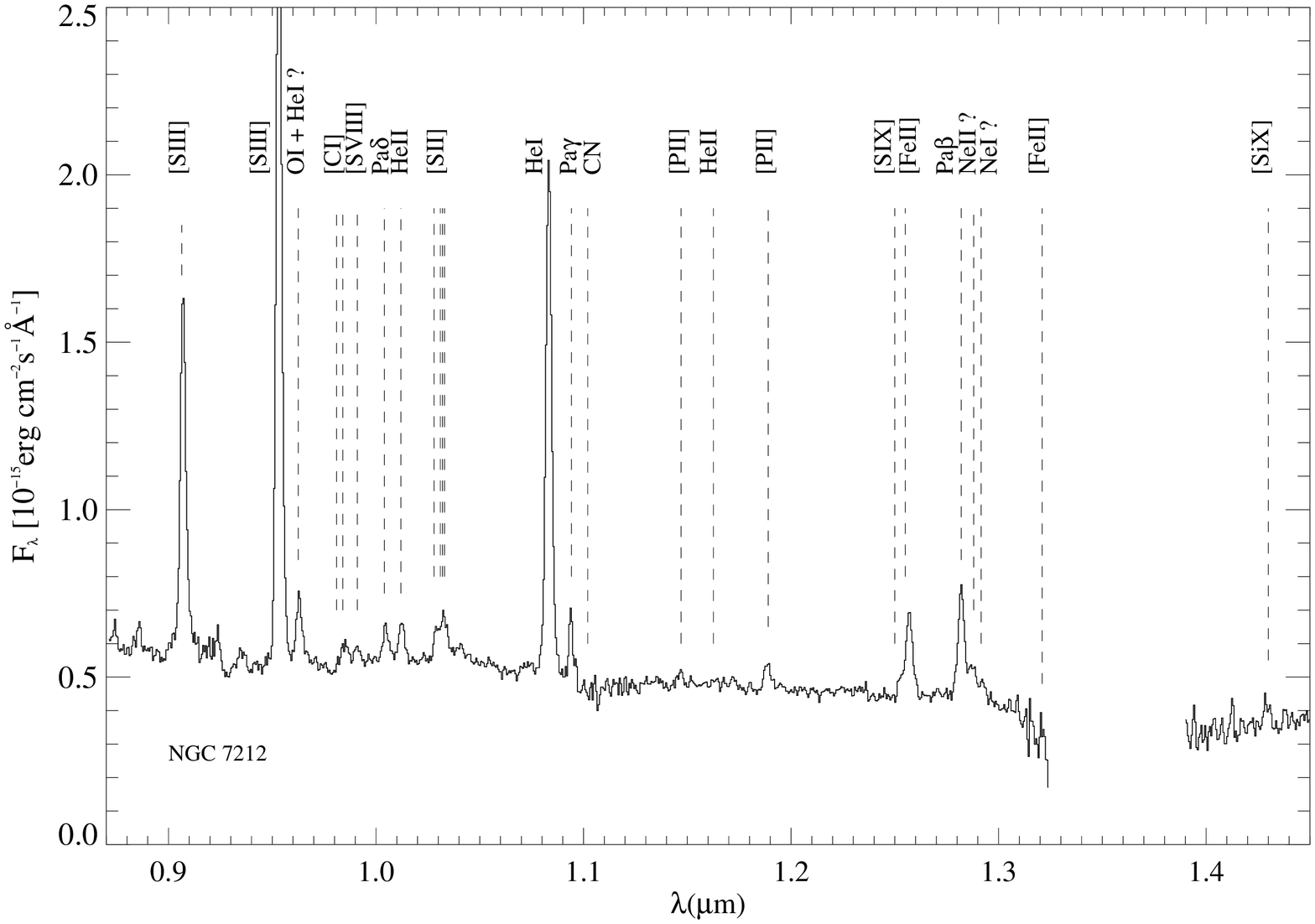}
\includegraphics[width=8cm,angle=90]{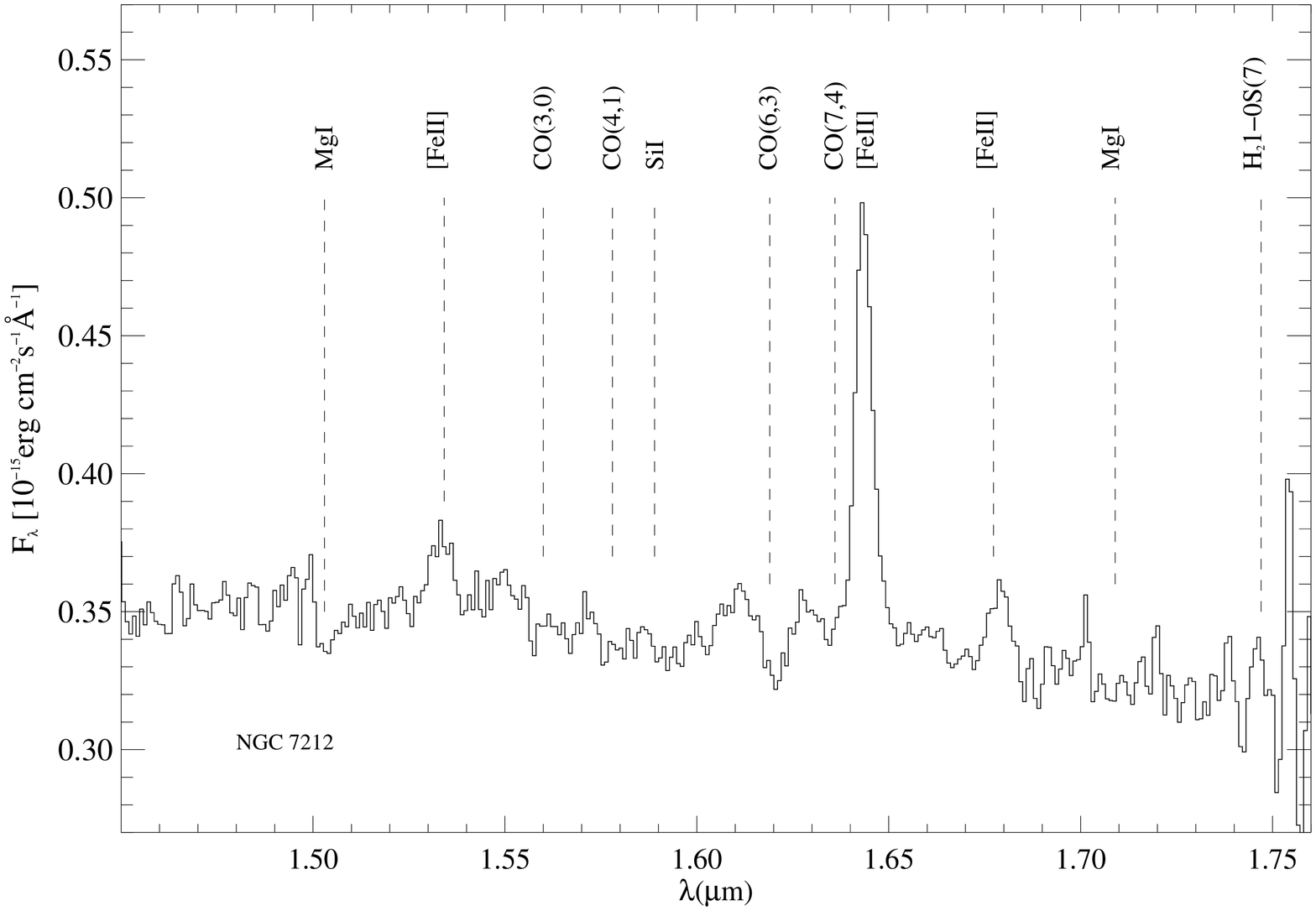}
\includegraphics[width=8cm,angle=90]{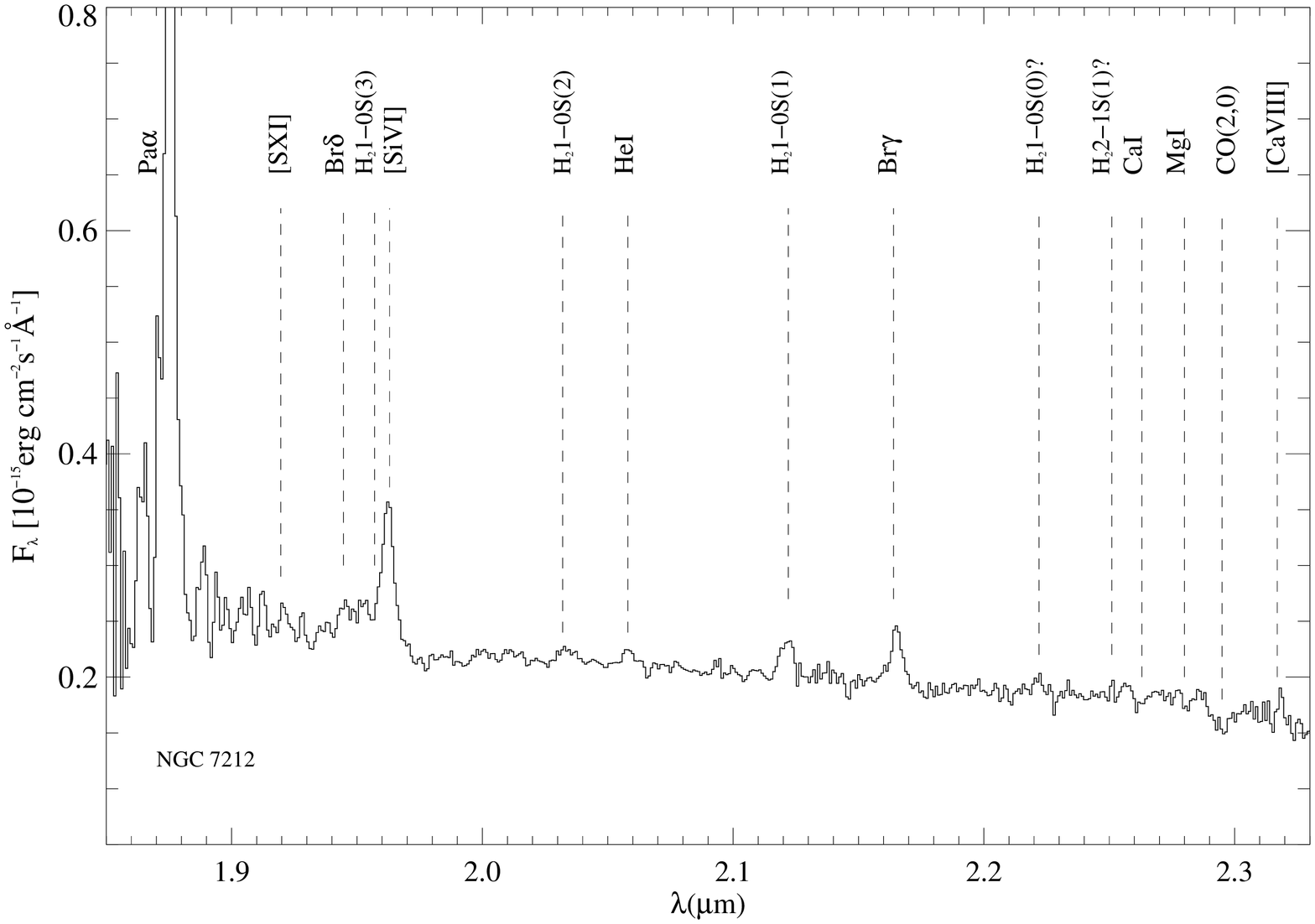}\par}
\figcaption{\footnotesize{Same as in Figure \ref{fi:m348}, but for NGC~7212.}
\label{fi:n7212}}
\end{figure}

\begin{figure}[!htp]
\centering
{\par
\includegraphics[width=8cm,angle=90]{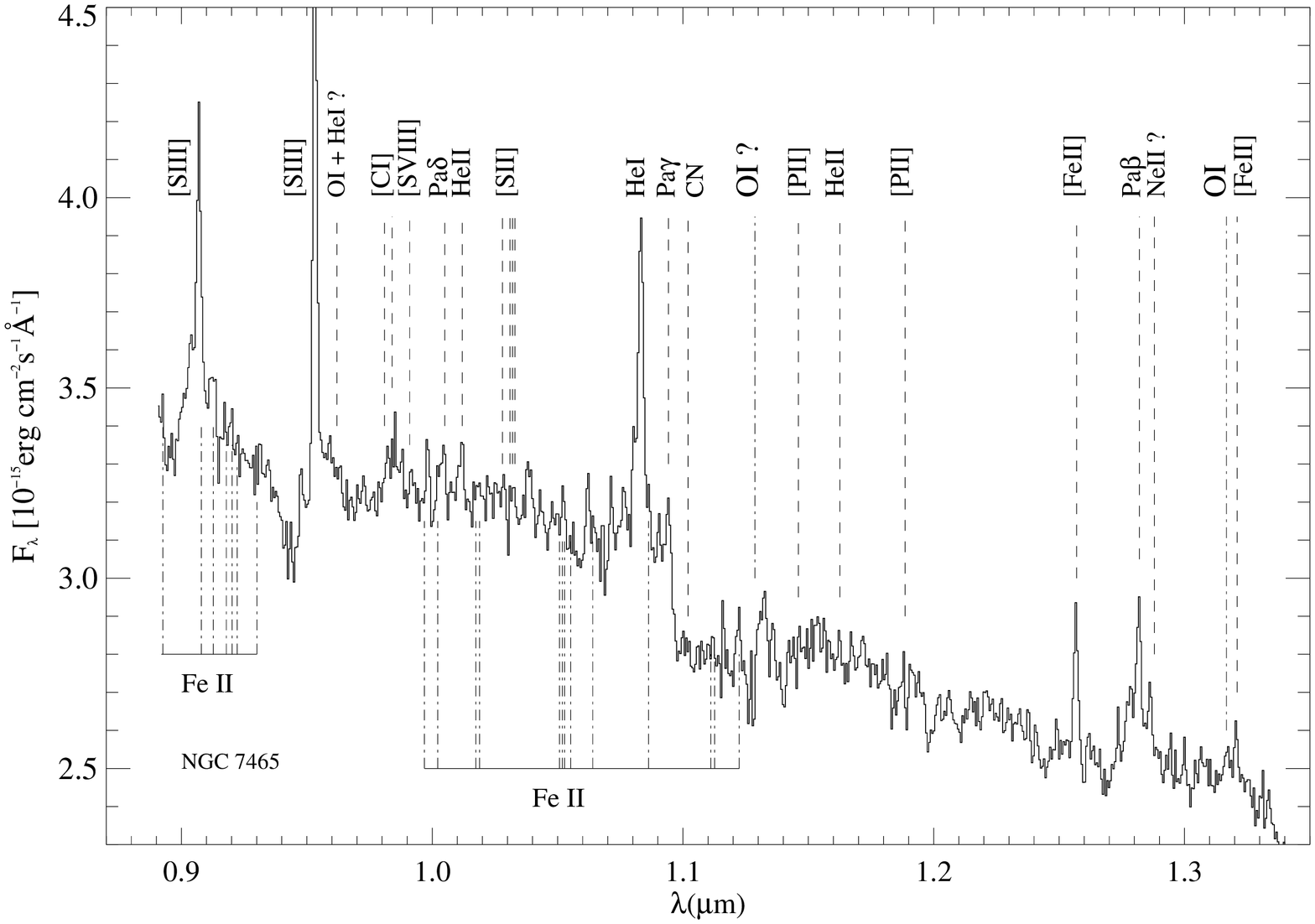}
\includegraphics[width=8cm,angle=90]{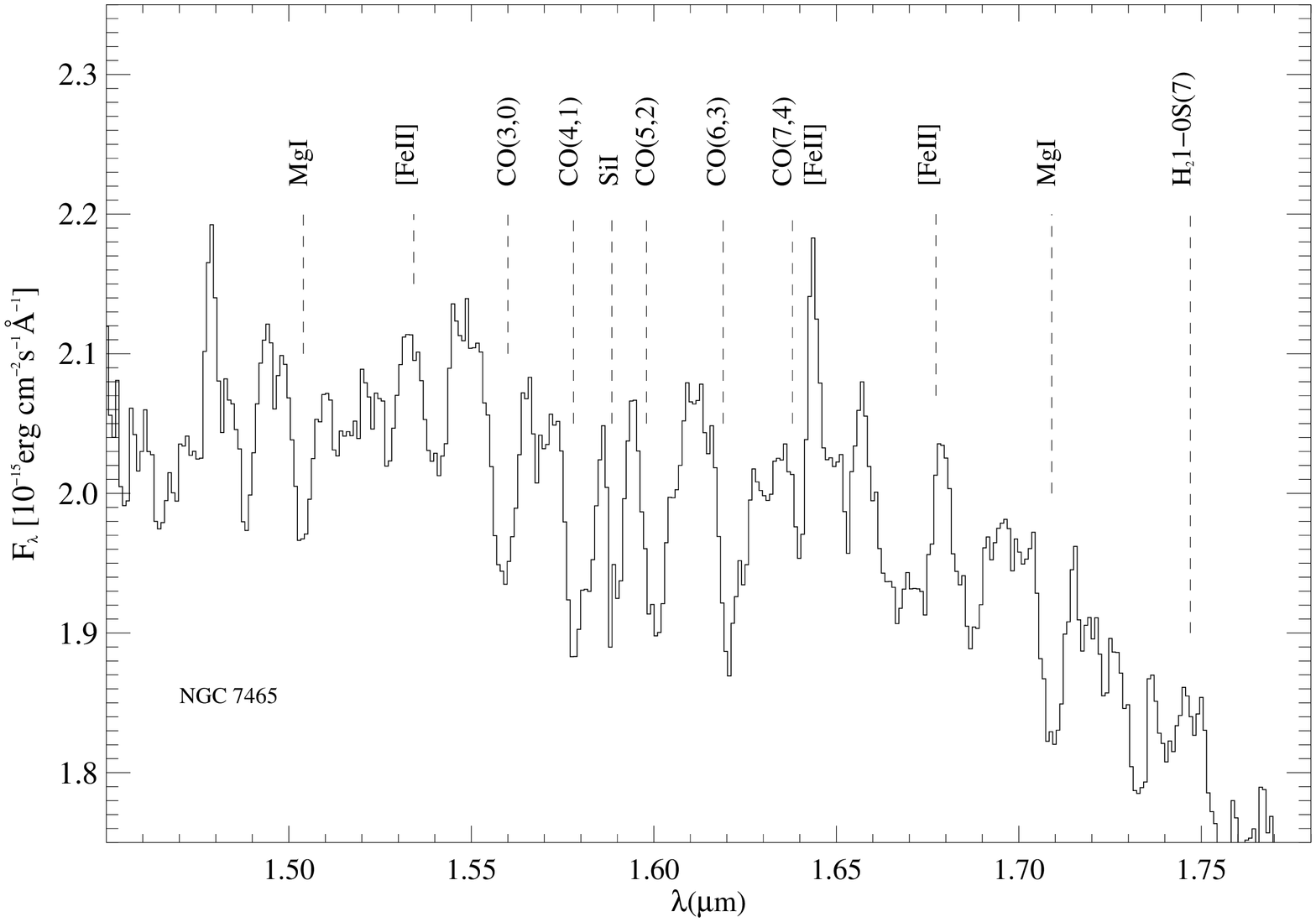}
\includegraphics[width=8cm,angle=90]{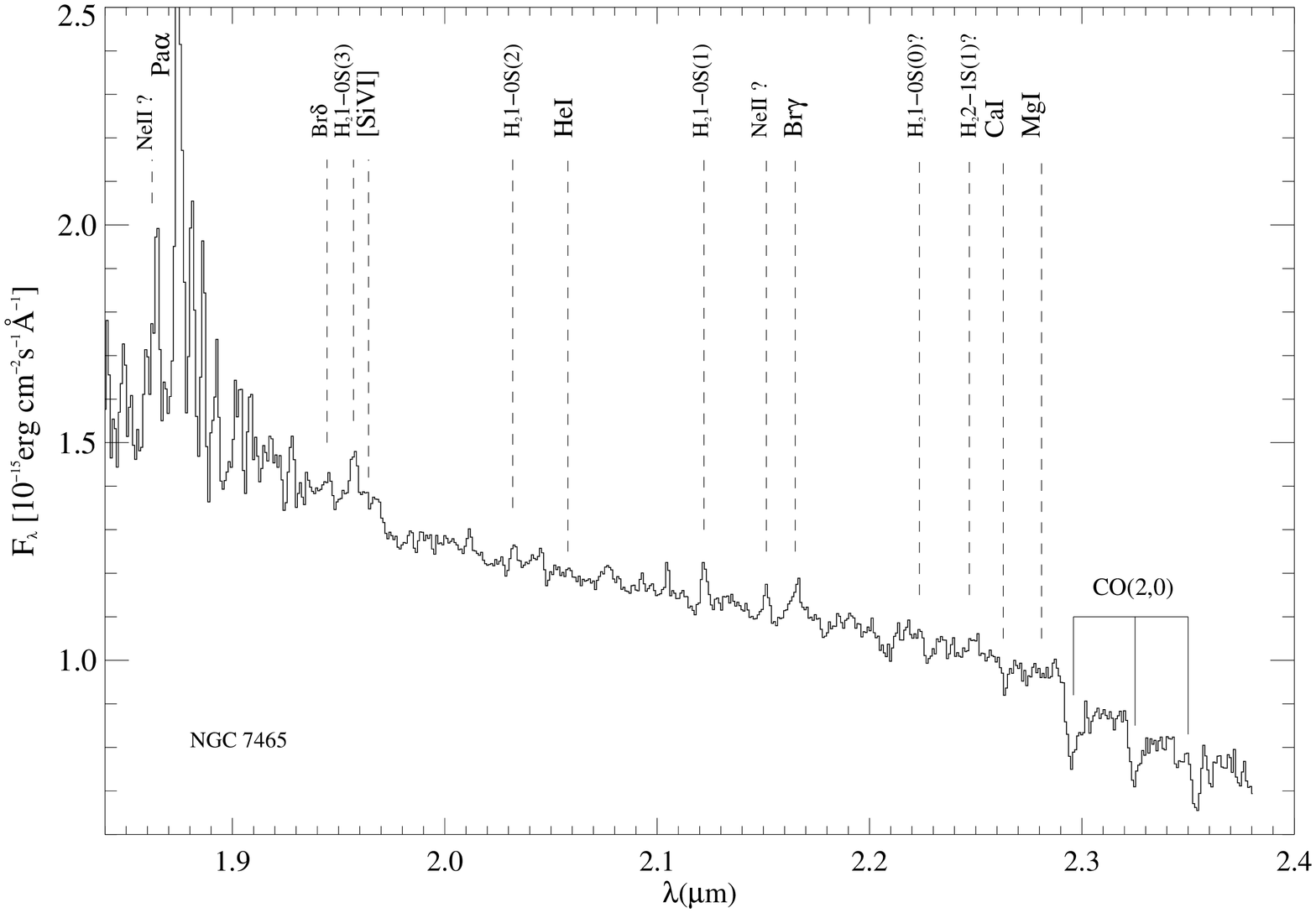}\par}
\figcaption{\footnotesize{Same as in Figure \ref{fi:m348}, but for NGC~7465. Note the broad pedestals of Pa$\beta$ and 
Br$\gamma$ as well as the presence of the permitted Fe II and O I lines.}
\label{fi:n7465}}
\end{figure}

The extracted nuclear spectra were then divided by their corresponding A0 spectra to remove telluric
contamination. 
A modified version of {\it Xtellcor} \citep{Vacca03} 
was used in this step. The resulting nuclear spectra
in the ZJ, H, and K bands are plotted in Figures \ref{fi:m348}, \ref{fi:m573}, \ref{fi:m1066}, \ref{fi:n7212}, 
and \ref{fi:n7465}, where the wavelength has been translated to the observer's rest frame.
Note that the absolute flux calibration is intended to be an approximation since 
the spectra of the comparison star is likely subjected to slit losses due to centering and tracking errors. 
Nevertheless, the agreement in the continuum flux level in the overlap region for the ZJ and HK
bands are quite good in spectral shape and absolute value.

\clearpage

\section{Results}

\subsection{The Emission Line spectra}
\label{emissionline}

The nuclear spectra are plagued of emission lines, as it can be seen in Figures 
\ref{fi:m348}, \ref{fi:m573}, \ref{fi:m1066}, \ref{fi:n7212}, and \ref{fi:n7465}. 
The features were measured by fitting a Gaussian using the Starlink program 
DIPSO. The resulting fluxes and FWHMs are reported in Table \ref{fluxes}.
In case of blended features two Gaussians have been employed to separate the different components. For example, 
the H$_{2}$ 1-0S(3) line is strongly blended with the coronal line [Si VI]$\lambda$1.963 in our resolution range.

The most prominent features in the nuclear spectra are [S III]$\lambda\lambda$0.907,0.953; 
He I$\lambda$1.083, [Fe II]$\lambda\lambda$1.257,1.644; Pa$\beta$; Pa$\alpha$; [Si VI]$\lambda$1.963;
and Br$\gamma$ (all the wavelengths are given in \micron). 
Several other lower intensity lines are detected in our spectra, namely, 
[C I]$\lambda$0.985; [S VIII]$\lambda$0.991; Pa$\delta$; He II$\lambda\lambda$1.012,1.163; 
[S II]$\lambda$1.029,1.032 (the reddest line is indeed a blend of three [S II] transitions); 
Pa$\gamma$; [P II]$\lambda$1.147,1.188, 
[Fe II]$\lambda\lambda\lambda\lambda$1.295,1.321,1.533,1.679,  He I$\lambda$2.058, and [Ca VIII]$\lambda$2.322.
The high-ionization transitions [S IX]$\lambda$1.252, [Si X]$\lambda$1.430, and [S XI]$\lambda$1.920 are 
also detected in the spectra of some objects. 
Broad wings in the Pa$\beta$ and Pa$\alpha$ profiles of Mrk~573 and in the Pa$\beta$ and Br$\gamma$ profiles of NGC~7465  
are clearly detected, together with the permitted O I$\lambda\lambda$1.128,1.317, Fe II 9200 \AA, and Fe II 1 \micron~lines.

A noticeable feature at 0.962 \micron~is clearly detected very close to the [S III]$\lambda$0.953 line 
in Mrk~348, Mrk~573, and NGC~7212 (see top panels of Figures \ref{fi:m348}, \ref{fi:m573}, and \ref{fi:n7212}). 
In NGC~7465 it seems to be blended with the [S III] line (see Figure \ref{fi:n7465}).
The detection of this line in Mrk~573 was firstly reported by \citet{Ramos08}, and it was tentatively identified as
O I (3D$_{0}$-3D) or He I (1P$_{0}$-1D). A blend of both transitions cannot be discarded, considering the large 
FWHM of the line (see Table \ref{fluxes}).

Emission from the H$_{2}$ molecule is clearly
detected in the galaxies of our sample, except for Mrk~573. The most prominent transitions are
H$_{2}$ 1-0S(3), H$_{2}$ 1-0S(2) and H$_{2}$ 1-0S(1). In the H band, the H$_{2}$ 1-0S(7) line is 
detected in most for the galaxies, whereas
both the H$_{2}$ 2-1S(1) and H$_{2}$ 1-0S(0) emission lines are
barely detected at  $\le$ 2$\sigma$ level in the majority of the galaxies, except in the case of Mrk~1066.

\newpage

\begin{landscape}
\begin{table}[ !ht ]
\centering
\tiny
\begin{tabular}{lccccccccccc}
\hline
\hline
Line Nucleus &\multicolumn{1}{c}{$\lambda$}&\multicolumn{2}{c}{Mrk~348}&\multicolumn{2}{c}{Mrk~573} &
\multicolumn{2}{c}{Mrk~1066}&\multicolumn{2}{c}{NGC~7212}&\multicolumn{2}{c}{NGC~7465} \\
&(\micron) & Flux & FWHM & Flux & FWHM & Flux & FWHM & Flux & FWHM & Flux & FWHM  \\
\hline
$[S III]$	& 0.907 &  63.4 $\pm$ 1.6  & 600 $\pm$ 20  &  116   $\pm$ 3   &  470 $\pm$ 20  &  157$\pm$4     & 300$\pm$20 & 35.2 $\pm$ 0.8 & 660 $\pm$ 30  & 21.5$\pm$2.6 & 750$\pm$170 \\
$[S III]$	& 0.953 & 141.3 $\pm$ 1.5  & 420 $\pm$ 10  &  279   $\pm$ 3   &  378 $\pm$ 8   &  429$\pm$4     & 340$\pm$10 & 82   $\pm$ 1   & 550 $\pm$ 10  & 32.3$\pm$1.9 & ...         \\
O I + He I ?    & 0.962 &   5.9 $\pm$ 0.9  & 920 $\pm$ 110 &  26.5  $\pm$ 1.7 &  680 $\pm$ 50  &  ...           & ...        &  5.6 $\pm$ 0.4 & 510 $\pm$ 60  & ...          &  ...        \\ 
$[C I]	$       & 0.982 &   2.7 $\pm$ 0.6  & 680 $\pm$ 190 &  2.9   $\pm$ 0.6 &  260 $\pm$ 90  &  11.3$\pm$2.4  & ...        &  0.3 $\pm$ 0.6 & ...           & 1.89$\pm$1.27& ...         \\ 
$[C I]	$       & 0.985 &   5.1 $\pm$ 0.4  & 420 $\pm$ 50  &  5.4   $\pm$ 0.8 &  250 $\pm$ 90  &  40.4$\pm$3.1  & 620$\pm$70 &  2.3 $\pm$ 0.6 & 730 $\pm$ 260 & 2.57$\pm$1.36& 400$\pm$400 \\  
$[S VIII] $     & 0.991 &   2.1 $\pm$ 0.4  & 160 $\pm$ 90  &  13.6  $\pm$ 0.8 &  200 $\pm$ 40  &  ...           & ...        &  1.6 $\pm$ 0.3 & 540 $\pm$ 180 & 3.1 $\pm$3.7 & 560$\pm$560 \\ 
Pa$\delta$      & 1.005 &   2.7 $\pm$ 0.4  & 260 $\pm$ 90  &  4.4   $\pm$ 0.5 &  ...	       &  23.5$\pm$1.4  & 30$\pm$30  &  3.1 $\pm$ 0.3 & 560 $\pm$ 100 & 2.8 $\pm$0.9 & 400$\pm$200 \\
He II	        & 1.012 &   4.6 $\pm$ 0.4  & 520 $\pm$ 70  &  21.3  $\pm$ 0.8 &  510 $\pm$ 30  &  13.5$\pm$1.7  & 560$\pm$90 &  3.7 $\pm$ 0.3 & 570 $\pm$ 80  & 2.8 $\pm$0.8 & 190$\pm$190 \\ 
$[S II]$	& 1.029 &   6   $\pm$ 1    & 500 $\pm$ 100 &  5.6   $\pm$ 0.7 &  210 $\pm$ 80  &  9.98$\pm$1.34 & ...        &  1.5 $\pm$ 0.5 & ...           & ...          & ...         \\
$[S II]$        & 1.032 &  16.4 $\pm$ 1.4  & 1100$\pm$ 100 &  9.4   $\pm$ 0.9 &  670 $\pm$ 90  &  30.6$\pm$1.8  & 650$\pm$60 &  6   $\pm$ 1   & 1100$\pm$ 190 & ...          & ...         \\
He I	        & 1.083 &  79   $\pm$ 2    & 540 $\pm$ 20  &  119   $\pm$ 2   &  370 $\pm$ 14  &  240$\pm$4     & 410$\pm$10 & 52.1 $\pm$ 0.6 & 580 $\pm$ 10  & 34.9$\pm$2.9 &1060$\pm$130 \\
Pa$\gamma$      & 1.094 &   7.1 $\pm$ 1.2  & 650 $\pm$ 150 &  16.6  $\pm$ 1.6 &  430 $\pm$ 60  &  59.4$\pm$3.7  & 240$\pm$40 &  4.8 $\pm$ 0.6 & ...           & 4.5 $\pm$1.4 & 220$\pm$180 \\
O I 		& 1.128 &   ...            & ...           &  2.6   $\pm$ 0.9 & ...	       &  ...           & ...  	     &  ... 	      & ... 	      & ...          & ...         \\  
$[P II]$        & 1.147 &   1.6 $\pm$ 0.3  & 640 $\pm$ 140 &  3.3   $\pm$ 0.7 & 340  $\pm$ 140 &  18.9$\pm$2.6  & 580$\pm$120&  0.8 $\pm$ 0.2 & ...           & ...          & ...         \\  
He II		& 1.163 &   0.7 $\pm$ 0.3  & ...           &  ...	      & ...	       &  3.81$\pm$1.79 & ...        &  ... 	      & ... 	      & ...          & ...         \\    
$[P II]$	& 1.188 &   2.9 $\pm$ 0.4  & 510 $\pm$ 120 &  3.4   $\pm$ 0.7 & ...	       &  33.6$\pm$2.9  & 620$\pm$90 &  2.8 $\pm$ 0.3 & 630 $\pm$ 90  & :1.6         & ...         \\  
$[S IX]$	& 1.252 &   2.9 $\pm$ 0.7  & 500 $\pm$ 100 &  8.5   $\pm$ 0.7 & 200  $\pm$ 40  &  ...           & ...  	     &  1.0 $\pm$ 0.2 & ...           & ...          & ...         \\ 
$[Fe II]$	& 1.257 &  12   $\pm$ 1    & 400 $\pm$ 40  &  17    $\pm$ 1   & 620  $\pm$ 50  &  92.2$\pm$1.9  & 360$\pm$10 & 10.2 $\pm$ 0.4 & 750 $\pm$ 40  & 8.2$\pm$0.7  & 200$\pm$50  \\ 
Pa$\beta$ narrow& 1.282 &  12.1 $\pm$ 0.6  & 340 $\pm$ 30  &  15.7  $\pm$ 1.3 & ...	       &  116.3$\pm$2.8 & 160$\pm$10 & 11.9 $\pm$ 0.5 & 580 $\pm$ 30  & 6.4$\pm$1.6  & 120$\pm$80  \\
Pa$\beta$ broad	& 1.282 &   ...            & ...           &  16    $\pm$ 3   & 1700 $\pm$ 400 &  ...           & ...  	     &  ... 	      & ... 	      & 13 $\pm$3    &2300$\pm$450 \\
Ne II ?      	& 1.287 &   5   $\pm$ 1    & 770 $\pm$ 140 &  10.3  $\pm$ 1.7 & 270  $\pm$ 50  &  15.7$\pm$2.6  & ...        &  4.6 $\pm$ 0.6 & 800 $\pm$ 150 & 2.2$\pm$0.8  & ...         \\ 
Ne I ?     	& 1.292 &   0.4 $\pm$ 0.2  & ...           &  8.6   $\pm$ 0.9 & 350  $\pm$ 60  &  8.55$\pm$2.55 & ...        &  1.9 $\pm$ 0.5 & 450 $\pm$ 180 & ...          & ...         \\
O I 		& 1.317 &   2.3 $\pm$ 0.9  & 650 $\pm$ 200 &  8     $\pm$ 2   & 320  $\pm$ 100 &  ...           & ...  	     &  ... 	      & ... 	      & 3  $\pm$ 1   & 540$\pm$260 \\
$[Fe II]$ 	& 1.321 &  10.8 $\pm$ 1.2  & 800 $\pm$ 100 &  13    $\pm$ 3   & 480  $\pm$ 160 &  33.5$\pm$4.9  & 600$\pm$100&  3.5 $\pm$ 1.4 & 300 $\pm$ 200 & 3.1$\pm$ 0.8 & ...         \\
$[Si X]$	& 1.430 &  ...             & ...           &  17.3  $\pm$ 2.8 & 330  $\pm$ 80  &  11.4$\pm$3.8  & 280$\pm$120&  2.2 $\pm$ 0.4 & 400 $\pm$ 100 & ...          & ...         \\
\hline	   						        	        			        																																																		  
$[Fe II]$	& 1.533 &   2.2 $\pm$ 0.3  & 800 $\pm$ 150 &  7.7   $\pm$ 1.4 & 1200 $\pm$ 200 &  11.7$\pm$3.4  & 580$\pm$220&  1.7 $\pm$ 0.3 & 970 $\pm$ 180 & 4.8$\pm$1.7  & 930$\pm$370 \\
$[Fe II]$	& 1.644 &  14.5 $\pm$ 0.6  & 660 $\pm$ 40  &  11.3  $\pm$ 1.8 &  510 $\pm$ 130 &  95$\pm$4      & 330$\pm$30 &  9.0 $\pm$ 0.3 & 720 $\pm$ 40  & 6.7$\pm$1.4  & 350$\pm$140 \\
$[Fe II]$	& 1.679 &   2.6 $\pm$ 0.6  & 640 $\pm$ 170 &  8.3   $\pm$ 2.4 & 1000 $\pm$ 300 &  16.99$\pm$4.43& 530$\pm$160&  2.0 $\pm$ 0.3 & 870 $\pm$ 110 & 6.8$\pm$0.6  & 790$\pm$80  \\
H$_{2}$ 1-0S(7)	& 1.747 &   1.7 $\pm$ 0.3  & 870 $\pm$ 160 &  3.2   $\pm$ 2.1 &  680 $\pm$ 470 &  17.05$\pm$4.14& 850$\pm$240&  1.04$\pm$0.43 & ...           & 4.4$\pm$1.1  &1070$\pm$250 \\
Ne II ?         & 1.866 &   ...            & ...           &  17.1  $\pm$ 2.8 &  ...	       &  ...           & ...        &  ... 	      & ... 	      & ...          & ...         \\   
Pa$\alpha$ narrow&1.875 &  35.5 $\pm$ 1.5  & 440 $\pm$ 30  &  37.2  $\pm$ 8.6 &  ...	       &  185$\pm$34    & 910$\pm$180& 37.1 $\pm$ 1.8 & 620 $\pm$ 50  &39.9$\pm$3.7  & 390$\pm$60  \\
Pa$\alpha$ broad& 1.875 &   ...		   & ...           &  70.9  $\pm$ 10.1& 1160 $\pm$ 210 &  ...           & ...        &  ... 	      & ... 	      & ...          & ...         \\   
$[S XI]$	& 1.920 &   ...            & ...           &   9.2  $\pm$ 2.1 &  440 $\pm$ 110 &  ...           & ...        &  0.7 $\pm$ 0.3 & ...           & ...          & ...         \\ 
Br$\delta$	& 1.944 &   3.3 $\pm$ 1.3  & 900 $\pm$ 300 &  ...	      &  ...	       &  11.8$\pm$2.6  & ...        &  0.5 $\pm$ 0.2 & ...           & ...          & ...         \\  
H$_{2}$ 1-0S(3)	& 1.957 &   3.4 $\pm$ 0.6  & 290 $\pm$ 120 &  6.6   $\pm$ 1.9 &  400 $\pm$ 200 &  36.1$\pm$4.7  & ...        &  ...           & ... 	      & 4.05$\pm$0.65& 350$\pm$80  \\
$[Si VI]$	& 1.963 &   9   $\pm$ 1    & 450 $\pm$ 50  &  27    $\pm$ 2   &  350 $\pm$  50 &  9.7$\pm$3.1   & ...        &  6.7 $\pm$ 0.4 & 610 $\pm$ 50  & 0.3$\pm$0.4  & ...         \\
H$_{2}$ 1-0S(2)	& 2.032 &   0.88$\pm$ 0.32 & 320 $\pm$ 220 &  :3.8	      &   ...	       &  13.8$\pm$1.6  & ...        &  0.8 $\pm$ 0.1 & 810 $\pm$ 160 & 1.4$\pm$0.3  & ...         \\
He I		& 2.060 &   ...            & ...           &  ...	      &   ...	       &  24.7$\pm$2.2  & 530$\pm$70 &  0.7 $\pm$ 0.1 & 470 $\pm$ 150 & ...          & ...         \\
H$_{2}$ 1-0S(1)	& 2.121 &   2.8 $\pm$ 0.3  & 140 $\pm$ 50  &  3.1   $\pm$ 1.3 &  190 $\pm$ 190 &  51.8$\pm$1.9  & 220$\pm$20 &  2.0 $\pm$ 0.2 & 600 $\pm$ 80  & 3.2$\pm$0.4  & 200$\pm$50  \\
Br$\gamma$ narrow&2.165 &   3.01$\pm$ 0.42 & 400 $\pm$ 100 &  3.5   $\pm$ 1.4 &  300 $\pm$ 200 &  44.3$\pm$1.8  & 100$\pm$20 &  2.8 $\pm$ 0.2 & 550 $\pm$ 70  & 3  $\pm$1    & 540$\pm$210 \\
Br$\gamma$ broad& 2.165 &   ... 	   & ...	   &  ...	      &   ...	       &  ...           & ...	     &  ... 	      & ... 	      &18.3$\pm$6.8  &2900$\pm$500 \\
H$_{2}$ 1-0S(0)	& 2.222 &   :0.6           & ...           &  :2.8	      &   ...	       &  16.4$\pm$2.5  & 260$\pm$70 &  0.6 $\pm$ 0.2 & 190 $\pm$ 190 & :1.6         & ...         \\ 
H$_{2}$ 2-1S(1)	& 2.248 &   :0.6	   & ...	   &  :2.8	      &   ...	       &  8.11$\pm$2.07 & ...	     &  :0.6 	      & ... 	      & :1.6         & ...         \\ 
$[Ca VIII]$     & 2.322 &   1.98$\pm$ 0.48 & 400 $\pm$ 100 &  ...	      &   ...	       &  ...           & ...	     &  0.7 $\pm$ 0.2 & ...           & ...          & ...         \\ 
\hline	   					 											        																																													  
\end{tabular}	
\caption{\scriptsize{Emission lines detected, central wavelength (rest frame), line fluxes (10$^{-15}$ erg~cm$^{-2}$~s$^{-1}$), 
and FWHM (in km~s$^{-1}$) corrected for instrumental broadening. Lines whose fluxes are reported and whose FWHM values are not are narrower than
the instrumental profile. For several faint transitions, the fluxes reported correspond to an upper limit of 2$\sigma$ (those beginning with :).}
\label{fluxes}}
\end{table}
\end{landscape}

\subsection{The Absorption Line Spectra}
\label{stars}

Various stellar absorption features populate the H and K nuclear spectra of our galaxies 
(see Figures \ref{fi:m348}, \ref{fi:m573}, \ref{fi:m1066}, \ref{fi:n7212} and \ref{fi:n7465}). The most
prominent ones are Mg I 1.50, CO 1.56, 1.58, Si I 1.59, CO 1.62, 1.64, and Mg I 1.71  (all
wavelengths in \micron) in the H band. These stellar features are named depending on 
their principal contributor, that depends on the spectral type. 
For example, for early M types, the Si I 1.59 absorption is due to silicon, 
whereas for very cool stars, OH dominates it \citep{Origlia93,dallier96}.
In the case of Mrk~573 and NGC~7465, also the CO 1.60 \micron~absorption band appears very deep.
The $^{12}$CO(2,0)~2.29 band is detected in all the galaxies, together with 
the Ca I 2.26 and Mg I 2.28 \micron~bands.
In the J range, it is noticeable the presence of the 1.1 \micron~CN band in the spectra of all the
observed galaxies, firstly reported by \citet{Riffel07} for Mrk~573 and Mrk~1066.

The equivalent widths (EWs) of the Si I 1.59, CO(6,3) 1.62, and CO(2,0) 2.29 \micron~bands are 
reported in Table \ref{absorption} for the five galaxies in our sample.
These  absorption line features were measured using the Starlink program 
DIPSO. 

\begin{table}[!ht]
\scriptsize
\centering
\begin{tabular}{cccccc}
\hline
\hline
Galaxy/Star class & \multicolumn{3}{c}{EW (\AA)} & H band dilution & K band dilution \\ 
 & Si I$\lambda$1.589 & CO(6,3)$\lambda$1.619 & CO(2,0)$\lambda$2.29 \\ \hline 
Mrk~348&  1.0$\pm$0.2 & 2.8$\pm$0.2 & 1.1$\pm$0.2 & 62-66\% (40\%) & 92\% (77\%)    \\ 
Mrk~573&  1.2$\pm$0.2 & 4.6$\pm$0.3 & 8.0$\pm$0.9 & 37-46\% (15\%) & $<$59\% (45\%) \\ 
Mrk~1066& 3.0$\pm$0.1 & 4.5$\pm$0.4 &12.7$\pm$0.7 &  ... (14\%)   & $<$8\% (48\%)  \\ 
NGC~7212& 1.4$\pm$0.2 & 5.4$\pm$0.4 & 5.9$\pm$0.9 & 26-37\% (0\%) & $<$69\% (20\%) \\ 
NGC~7465& 1.8$\pm$0.2 & 5.0$\pm$0.3 &10.5$\pm$1.5 & 32-38\% (6\%) & 24\% (32\%)   \\ 
K5 III& 	  2.8         & 7.8 	    & 14.1        & ...	  & ...	  \\ 
M3 III&    2.9 	      & 8.1         & 15.3        & ...	  & ...	  \\ 
K4 I  &    2.2         & 7.3         &  ...	  & ...	  & ...	  \\ 
M1 I  &    2.8         & 8.6         & 19.4	  & ...	  & ...	  \\ 
\hline
\end{tabular}
\caption{\footnotesize{Absorption line EWs in the H and K bands for the five galaxies 
in our sample and for the two giant and supergiant stellar types considered. 
The last two columns correspond to the starlight dilution in the H and K bands. Values within
parenthesis correspond to the dilution fraction obtained from the continuum fitting  described in Section 
\ref{sc:fitcont}.}
\label{absorption}}
\end{table}

\section{Discussion}

\subsection{The Underlying Stellar Population}

\begin{figure}[!h]
\centering
\includegraphics[width=9cm,angle=90]{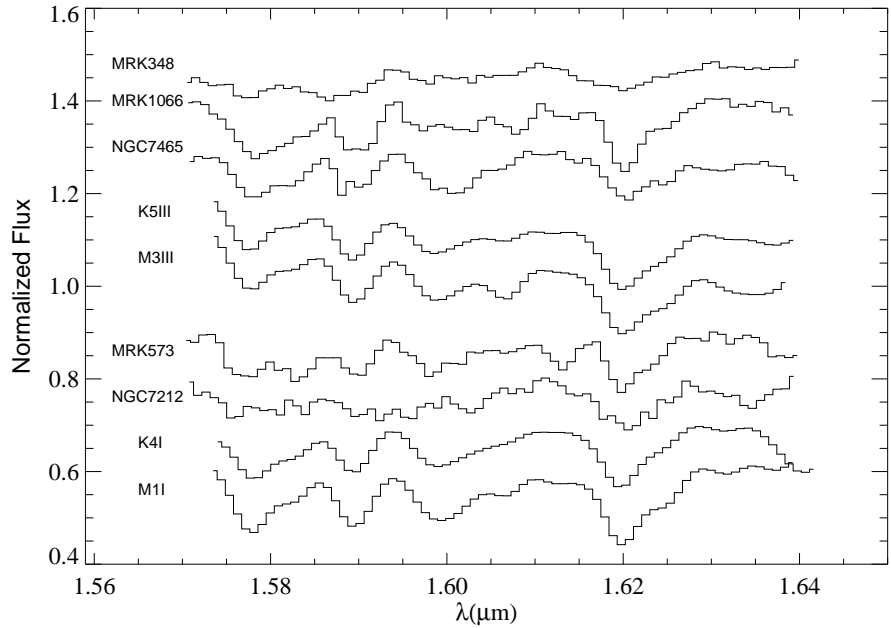}
\caption{\footnotesize{H band spectra of Mrk~348, Mrk~1066, NGC~7465, 
and two giant stars (K5 III and M3 III types) at the top, and the same for Mrk~573, 
NGC~7212, and two supergiant stars (K4 I and M1 I) at the
bottom. The spectra of stars have been convolved and rebinned in order to obtain
the same resolution as those of the galaxy spectra. All spectra are normalized
and shifted vertically for clarity.}}
\label{hstar}
\end{figure}

To identify the spectral types that produce the absorption features in 
our galaxies to a first approximation, we compare our H band spectra with digitally
avalaible stellar templates from \citet{dallier96} (see Figure \ref{hstar}).
After visual inspection, it can be noticed that the nuclear spectra resemble those of late-type stars\footnote{In this work
we always refer the emission towards the nucleus as the nuclear stellar population although we cannot discriminate between
nuclear-located and foreground stellar emission, because of the projection effects along the LOS.}. 
In order to establish which are the stellar populations dominant in each galaxy, measurements of the EWs of 
several absorption features (Si I 1.59 \micron, CO(6,3) 1.62 \micron, and CO(2,0) 2.29 \micron~bands. See Table \ref{absorption}) 
are compared with those derived from stellar spectra.  
The intrinsic stellar features reported in Table \ref{absorption}
have been computed after convolving by a Gaussian kernel, in order to match the resolution of our 
galaxy spectra. 

In particular, the ratio of the EW of the 1.62 \micron~CO(6,3) 
feature to that of the 1.59 \micron~Si I is a good temperature indicator for late-type stars
\citep{Origlia93}.
In this case, dilution and reddening effects are cancelled out, due to the closeness of the
features in wavelength \citep{Origlia93,forster00,ivanov04}.
The EW ratio of the CO 1.62 feature to Si I 1.59~\micron\ corresponds to late spectral 
types for all galaxies in our sample (see Table \ref{absorption}). Late-type giants 
are dominating the near-infrared spectra of 
Mrk~348, Mrk~1066, and NGC~7465, which 
is consistent with the results obtained by \citet{Ramos06} for
Mrk~78 and those by \citet{oliva95} for a variety of AGN. 
The values of the CO~1.62 to SiI~1.59 ratio for Mrk~573 and NGC~7212 are larger, 
closer to late-type supergiants. 
Nevertheless, the use of this ratio to discriminate between old stellar population (red giants) 
and recent starburst (red supergiants) is not a reliable diagnostic \citep{oliva95}. 
A lower limit of $\sim 100$~Myr can be estimated for the age of the nuclear stellar population
of the studied galaxies, since intermediate-to-low-mass giants dominate 
the infrared spectrum \citep{Renzini86}.

As it is mentioned in Section \ref{stars}, we detect the 1.1 \micron~band in the J range of all 
our nuclear spectra.
A prominent CN band indicates the existence of thermal-pulsing AGB stars with
ages $\sim$ 1 Gyr \citep{Maraston05}, and it is associated with the presence of bright carbon 
stars \citep{Riffel07}. According to population synthesis models this feature appears very deep 
for ages $\sim 0.3-1$~Gyr, becoming diluted for older populations.

The stellar features may be substantially diluted by the contribution of the non-stellar emission.
Thus, the dilution fraction (D) can be estimated from the expression 
$D = 1 - \mathrm{EW}_{obs}/\mathrm{EW}_{int}$, being $\mathrm{EW}_{obs}$ and $\mathrm{EW}_{int}$
the observed and intrinsic EWs of a given absorption feature
\citep{oliva95}, assuming a single stellar population. 
Dilution fractions in the H and K bands have been computed using 
the CO~1.62~\micron\ and  CO~2.29~\micron~features, respectively.  
For giant stars, digitally available spectra from \citet{wallace97} are used.
The resulting dilution factors are given in Table \ref{absorption}, together with
those derived from the fits to stellar templates plus Blackbody (BB) component described below in 
Section \ref{sc:fitcont}.

Summarizing, the stellar population present in the nuclear region of
our five galaxies is dominated by late-type giants, although a low fraction of 
low-mass supergiants could be also present. The average age of these stellar populations 
must be between 100 Myr and 1 Gyr, according to the observed stellar features.

\subsection{The Continuum Spectra}
\label{sc:fitcont}

\begin{figure*}[!ht]
\centering
\includegraphics[width=10cm,angle=90]{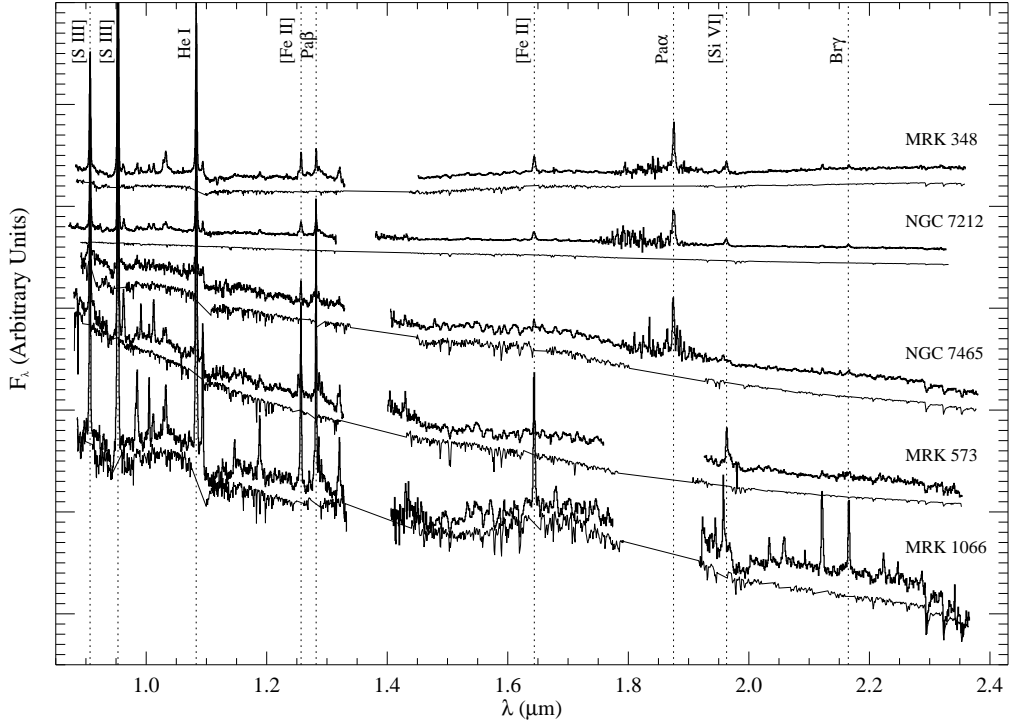}
\figcaption{\footnotesize{Normalized nuclear spectra  ordered according to their spectral shapes
from the flattest continuum (top) to the steepest one (bottom). The fitted templates have been shifted vertically for clarity. 
The wavelength has been translated to the observer's rest frame. Most prominent emission lines are labelled.
See Figures \ref{fit348}, \ref{fit573}, \ref{fit1066}, \ref{fit7212}, and \ref{fit7465} for color versions of the individual fits including residuals.}
\label{sed}}
\end{figure*}

The near-infrared continua of the five galaxies in our sample are shown in Figure \ref{sed}. 
The normalized spectra are ordered from the flattest continuum (Mrk~348) to the steepest (Mrk~1066). 
The continuum of Mrk~348 presents a peculiarity, compared with the rest of the galaxies.
There is a red excess, similar to the one observed in the spectrum of Mrk~1239 \citep{RodriguezArdila06}. 
These authors suggested the presence of a BB of T$\simeq 1200$~K
that would be dominating the emission in the L and M bands.

Given the multitude of stellar absorption features present in the nuclear spectra of our galaxies, it 
is reasonable to think that a major continuum contributor is the stellar population.  
In addition, it is expected that in the J band the emission from the active nucleus (that is the main contribution 
to the ultraviolet and optical ranges in Type-1 AGN) does not longer dominate \citep{Kishimoto05,Riffel06}.
The reprocessed nuclear emission coming from hot dust emerges in the K band 
as an important source of continuum \citep{RodriguezArdila06,Glikman06}. It seems clear that 
the best way of modelling the continuum of our galaxies must be based on a composition of stellar templates 
plus a non-stellar continuum which is a rising contribution at K band (BB).  
Indeed, the continuum spectra of our galaxies do not display a clear break at 1.1 \micron, 
contrary to what is seen in the spectra of Type-1 Seyferts, similarly to the 
findings reported by \citet{Riffel06}  for a large sample of AGN. This fact supports that 
a power-law component (associated to the active nucleus) is not relevant in this spectral range for Type-2 objects.


The continua have been fitted with stellar templates from the IRTF 
stellar library\footnote{Available in electronic form at http://irtfweb.ifa.hawaii.edu/$\sim$spex/WebLibrary/} \citep{Cushing05,Rayner09}, 
plus a BB component (see Figure \ref{sed} and Appendix \ref{sc:indiv}). The relative contributions
of the stellar template and BB component are allowed to vary freely.   
In addition, the stellar templates can suffer of a moderate amount of extinction, which is also a 
free parameter in the fitting process. 

The stellar templates that produce the best fitting model to the continuum range from spectral 
types K3 to M1, including dwarf and giant luminosity classes (see Appendix \ref{sc:indiv} for details on individual objects). 
In some cases the CO(2,0)~2.23 \micron\ bandhead and the depression at 1.1~\micron\ cannot be reproduced by the 
best fitting template, suggesting the additional contribution of a population of giant or 
supergiant stars, depending on the primary template, to match the details of the absorption features. 
The required extinction of the fitted stellar templates is negligible in all cases except in Mrk~348 and Mrk~1066,
for which is about E(B-V)=0.12.  In the case of Mrk~1066 this value is compatible with Galactic extinction (see Table \ref{info}).
A non-negligible contribution of the non-stellar BB component at the
K band is found for all the galaxies. 
Similarly, \citet{Glikman06} showed that a BB component is almost ubiquitous 
in the near-infrared spectrum of radio quiet AGN. 
The dilution fraction of the stellar spectra ranges from the lowest values of 
20\% in NGC~7212 and 32\% in NGC~7465, to the highest value of 77\% in Mrk~348 (see Table \ref{absorption}).

\subsection{Extinction towards the nucleus}

The near-infrared spectral region offers new and more reliable reddening diagnostics, compared to the optical 
range. Few ratios of hydrogen recombination lines can be used as reddening indicators, together 
with  suitable pairs of forbidden lines. 

In order to estimate the obscuration of the nuclear region of our galaxies, we have 
employed the Pa$\gamma$ to Pa$\beta$ and the [Fe II]$\lambda$1.257 to [Fe II]$\lambda$1.644 line ratios. 
The intrinsic Pa$\gamma$/Pa$\beta$ ratio depends on the density and temperature of the line-emitting gas 
\citep{Hummer87}. 
We have adopted theoretical values corresponding to T$_{e}$ = 10,000 K and N$_{e}$ = 10$^{4}$
cm$^{-3}$. In those cases where a broad component appears in Pa$\beta$ (Mrk~573 and NGC~7465) the line ratio 
has been computed using only the narrow component. The Pa$\alpha$ line was not used because it is subject 
to strong telluric absorption. 
In contrast, the [Fe II]1.257/1.644 line ratio seems to be a reliable indicator given the fact that it is less 
sensitive to the line-emitting gas density or temperature. The intrinsic value has been taken from 
\citet{Bautista98}.

\begin{figure}[!ht]
\centering
\includegraphics[width=8cm,angle=90]{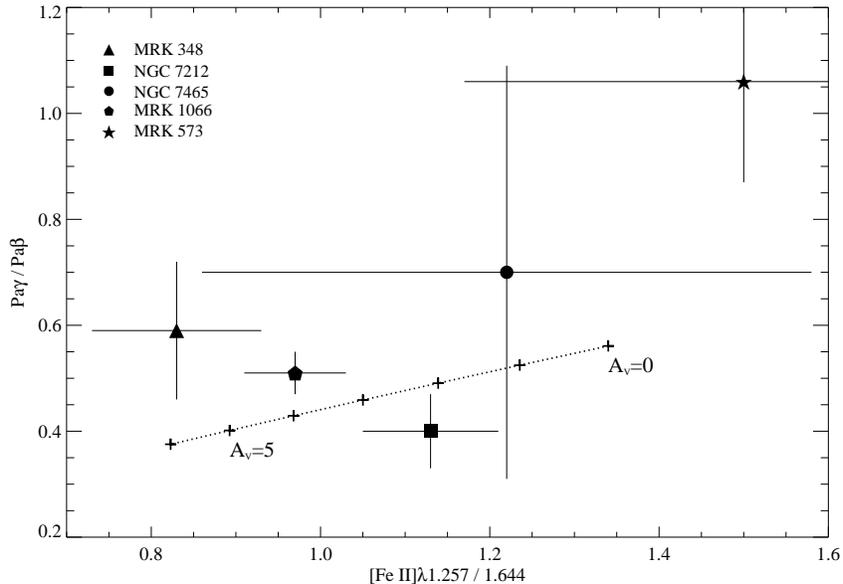}
\figcaption{\footnotesize{Pa$\gamma$/Pa$\beta$ versus [Fe II]$\lambda$1.257/1.644 for the five galaxies
in our sample. The theoretical sequence of A$_{V}$ values is represented with
the dashed line, with crosses marking increments of 1 mag.}
\label{reddening}}
\end{figure}

From the above line ratios we have determined the optical extinction (A$_{V}$) by using the Draine's parametrization,
A$_{\lambda}~\alpha~\lambda^{-1.75}$  \citep{Draine89}. 
The results are reported in Table \ref{extinction} and a comparison of the reddening estimations from both 
ratios is presented in Figure \ref{reddening}. As it can be seen in this figure there are discrepancies between
the two reddening estimations, a fact already discussed in \citet{Riffel06}.  
The intrinsic ratio of the recombination lines is very likely affected by high density and radiation transfer effects.  
Assuming that the [Fe II]1.257/1.644 line ratio is a more reliable indicator, an extinction sequence can be established 
from little reddened nuclei as Mrk 573 and NGC 7465 to the most reddened one, Mrk~348 ($A_V \sim 6$ mag), 
passing through NGC~7212 ($A_V \sim 2$ mag) and Mrk~1066  ($A_V \sim 4$ mag).  
It is noticeable that the two nuclei where broad emission lines have been detected present the less 
extinct NLRs. 
It is worth to remember that the extinction measured using this combination of narrow lines provides only an indication of 
the material located between us and the NLR. However, the innermost nucleus of the galaxy and the BLR may be hidden by
a larger amount of material. In particular, the non-detection of broad wings (if a hidden BLR is already present) in the 
recombination lines implies nuclear $A_V > 10$ mag  \citep{Veilleux97}. 

\begin{table}[ !ht ]
\centering
\scriptsize
\begin{tabular}{lccccccccccc}
\hline
\hline
Line Ratio & Theor. & \multicolumn{2}{c}{Mrk~348}&\multicolumn{2}{c}{Mrk~573} & 
\multicolumn{2}{c}{Mrk~1066}&\multicolumn{2}{c}{NGC~7212}&\multicolumn{2}{c}{NGC~7465} \\
& & Obs & A$_{V}$ &  Obs & A$_{V}$ &  Obs & A$_{V}$ & Obs & A$_{V}$  & Obs & A$_{V}$ \\
\hline
$[Fe II]$1.26/1.64 & 1.34 & 0.83$\pm$0.10 & 5.89 & 1.50$\pm$0.33 & ... & 0.97$\pm$0.06 & 3.91 & 1.13$\pm$0.08 & 1.97 & 1.22$\pm$0.36 & 1.10 \\ 
Pa$\gamma$/Pa$\beta$ & 0.56 & 0.59$\pm$0.13 & ...  & 1.06$\pm$0.19 & ... & 0.51$\pm$0.04 & 1.42 & 0.40$\pm$0.07 & 5.05 & 0.70$\pm$0.39 & ...  \\
\hline  																																											   
\end{tabular}  
\caption{\footnotesize{Theoretical values for the [Fe II]1.257/1.644 ratio
\citep{Bautista98} and for the Pa$\gamma$/Pa$\beta$ ratio \citep{Hummer87}. Measured values of these ratios from
our spectra, and optical extinctions calculated using Draine's parametrization are also reported.} 
\label{extinction}}
\end{table}

\subsection{Coronal Lines}

The simultaneous observation of very high- and low-ionization lines in the spectra of our 
Seyfert galaxies implies that a wide variety of physical conditions must coexist in the 
nuclear region of these objects. This constitutes a characteristic feature of gas photoionized
by a radiation source extending from the UV through the X-ray regime, typical of extreme
energetic environments such as AGN \citep{Prieto00}. 

In particular, the detection of strong coronal lines (IP $>$ 100 eV) such as  
[S VIII]$\lambda$0.991, [S IX]$\lambda$1.252, [Si X]$\lambda$1.430\footnote{The [Si X]$\lambda$1.430 
line is inmersed in a telluric absorption band due to the redshift of the galaxies considered, 
making difficult to detect it except under good atmospheric transmission.}, [Si VI]$\lambda$1.963, 
and [S XI]$\lambda$1.920
in the spectra of some of our galaxies is a confirmation of their AGN nature. 
These transitions are strong in Mrk~348, Mrk~573, and NGC~7212 
(see Figures \ref{fi:m348}, \ref{fi:m573} and \ref{fi:n7212}). 
These lines can be produced neither in starburst regions nor in low-luminosity AGN (LINERs) due to their high
ionization potentials. Indeed, coronal lines can only exist very close to the ionization source, 
making them unique tracers of AGN activity and energetics. 
For the other two galaxies, Mrk~1066 and NGC~7465, these lines
are extremely faint or undetected (see Figures \ref{fi:m1066} and \ref{fi:n7465}), indicating that their
nuclei are likely low-luminosity AGN (for a more detailed discussion of the individual properties of these objects 
see Appendices \ref{sc:m1066} and \ref{sc:n7465}). 

\subsection{Photoionization versus Shock Excitation}
\label{shocks}

It is well known that strong [Fe II] emission is indicative of shock-excited gas as occurs in the 
filaments of supernova remnants, in contrast to the weak [Fe II] emission characteristic of photoionized
gas as observed in H II regions. 
In AGN, strong [Fe II] emission is also common, although 
there is still some controversy about the dominant mechanism responsible for such emission. 
Several processes may contribute to the production of the [Fe II] lines: (1) photoionization by 
extreme UV to soft X-ray radiation from the central source, producing large  
partially-ionized regions in the NLR clouds; (2) interaction of radio jets with the surrounding medium, 
inducing shocks and hence partially ionized cooling tails; 
and (3) fast shocks associated with supernova remnants present in starburst regions. 
The [Fe II]$\lambda$1.257/Pa$\beta$ line ratio (or equivalently [Fe II]$\lambda$1.644/Pa$\alpha$) is 
very useful for distinguishing between a stellar or non-stellar origin of the 
[Fe II] emission. These line ratio increases from H II regions (photoionization by hot stars) 
to supernova remnants (shock excitation), passing through starburst and active galaxies 
\citep{Alonso97,Rodriguez04}.
Galaxies exhibiting [Fe II]$\lambda$1.257/Pa$\beta$ ratios lower han 0.3 are usually classified as 
starbursts, and those with values larger than 2 as LINERs.  
Seyfert galaxies are contained in the range 0.4-2 \citep{Larkin98,Rodriguez04}.

\begin{figure}[!ht]
\centering
\includegraphics[width=8cm,angle=90]{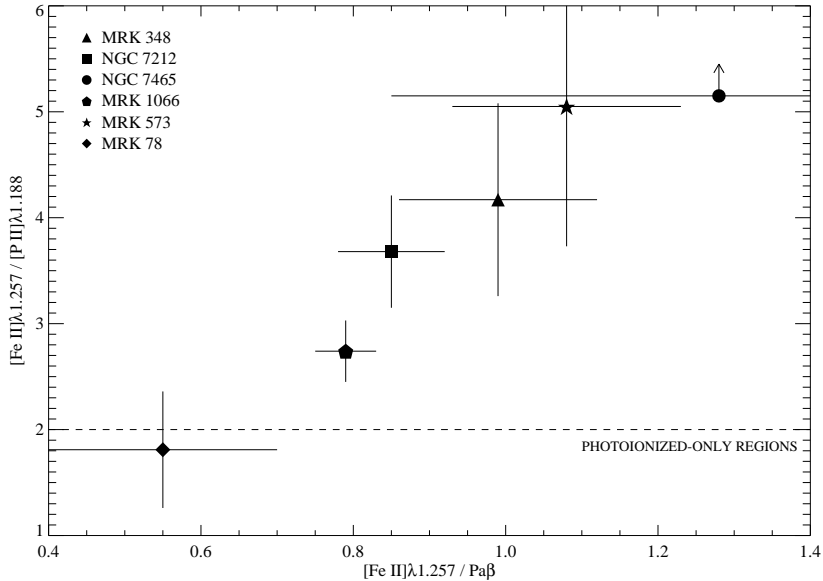}
\figcaption{\footnotesize{[Fe II]$\lambda$1.257/[P II]$\lambda$1.188 versus [Fe II]$\lambda$1.257/Pa$\beta$ 
for the five galaxies in our sample plus Mrk~78 (nuclear values taken from \citealt{Ramos06}). 
Dashed line represents the upper limit for pure photoionized gas.}
\label{iron}}
\end{figure}

The above mentioned line ratios for the five galaxies in our sample, plus the nuclear value of 
Mrk~78, taken from \citet{Ramos06}, are plotted in Figure \ref{iron}.  
In the cases of Mrk~573 and NGC~7465, the narrow-component of Pa$\beta$ has been employed 
to compute the ratio, since both galaxies have also broad components of the recombination lines. 
For all the galaxies, the [Fe II]$\lambda$1.257/Pa$\beta$ nuclear ratio is within the 
expected range for Seyfert galaxies, although NGC~7465 approaches to the LINER regime.

Another emission line useful for discriminating between the different excitation mechanisms acting
 in the NLR is [P II]$\lambda$1.188, which is detected in most of our spectra, together with the weaker [P II]$\lambda$1.147 transition.
In particular, [P II]$\lambda$1.188 
is very useful when compared with the [Fe II]$\lambda$1.257 line \citep{Oliva01}. Both lines are produced 
in partially ionized regions, having similar critical densities and excitation temperatures. 
However, iron is a well-known refractory species and is 
strongly depleted in dust grains, whereas phosphorus is not. 
Photoionization alone is unable to destroy the tough iron-based grains 
that are easily sputtered by shocks, passing to gaseous phase. Thus, the [Fe II]/[P II] 
ratio is high ($\ga$ 20) in fast shock-excited regions, and low ($\la$ 2) in pure photoionized 
regions \citep{Oliva01}. Values of this ratio for our five Seyfert galaxies and Mrk~78 are reported in Figure \ref{iron}.
For all the sources, except for the nucleus of Mrk~78\footnote{The [Fe II] emission increases from the nucleus towards the outer regions of 
Mrk~78, approaching the expected values for pure shock excitation, as already mentioned in Section \ref{intro} \citep{Ramos06}.}, 
the [Fe II]/[P II] ratio is larger than the predicted value for 
pure photoionized regions, indicating that interaction with the radio emission likely increases 
the proportion of iron in the partially ionized regions of these galaxies. This is specially critical in
Mrk~573 and NGC~7465, although they are still far from pure shock excitation. 

\subsection{Comparison with Photoionization Models}
\label{photoionization}

	We employed the photoionization code \textsc{Cloudy} (calculations were performed with version C07.02.01 of
\textsc{Cloudy}, last described by \citealt{Ferland98}) trying to reproduce line ratios measured from our nuclear spectra.
We have computed a grid of models based on photoionization by a given continuum and the expected 
physical conditions of 
the NLR. We used as ionizing continuum the so-called {\it Table agn} provided in \textsc{Cloudy}, that is very 
similar to that of typical radio quiet active galaxies (presented in \citealt{Mathews87}). 
A plane-parallel geometry was employed, metallicity equal to 0.3 solar, and grains with similar properties to
those of the Orion Nebula. 
Two different values of hydrogen density were explored, namely: n$_{H}$ = 10$^{3}$ and $10^{5}~\mathrm{cm}^{-3}$. 
The ionization parameter $U$, defined as $U = Q_{H}/(4 \pi d^{2} n_{H} c)$ was varied between 10$^{-4}$ and 10.

\begin{figure}[!ht]
\centering
\includegraphics[width=10cm,angle=90]{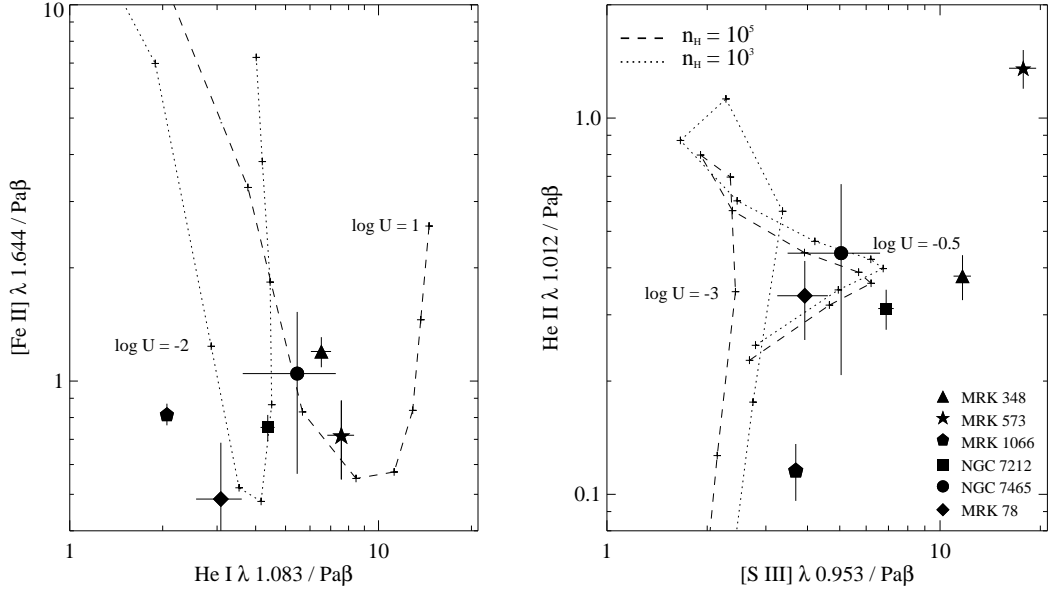}
\figcaption{\footnotesize{Diagnostic diagrams computed with \textsc{Cloudy}. Models are calculated using the  {\it Table agn} continuum, 
two diferent values of the hydrogen density (dotted and dashed lines) and a sequence of ionization parameters. 
The marks along the log $U$ sequence are separated by a step of 0.5.
Measurements from our spectra are overplotted with different symbols for each galaxy.}
\label{cloudy1}}
\end{figure}

Line ratio diagrams are shown in Figure \ref{cloudy1} for the following combination of lines: 
[Fe II]$\lambda$1.644 / Pa$\beta$ versus He I$\lambda$1.083 / Pa$\beta$
and He II$\lambda$1.012 / Pa$\beta$ versus [S III]$\lambda$0.953 / Pa$\beta$, respectively.
The dotted and dashed lines represent the computed line ratios for the two hydrogen densities considered, varying 
the ionization parameter from log $U$ = 1 to -4 with a step of 0.5.
Values of the observed line ratios from Table \ref{fluxes}, corresponding to the five galaxies considered in this work
are overplotted for comparison with our calculations made with \textsc{Cloudy}.  
We have also included the nuclear Mrk~78 measurements in these diagnostic diagrams \citep{Ramos06} for comparison with the rest of
the galaxies.

The diagnostic diagram involving the [Fe II] and He I lines predicts values of the ionization parameter larger than 
log $U$ $\sim$ -2 for all the galaxies considered, while the one including He II and [S III] is less restrictive. 
Actually, the computed models do not predict a simple behaviour for the considered line ratios, making difficult to determine any
sound value of the ionization parameter capable of explaining simultaneously our measurements for most of the galaxies. More sophisticated 
modelling is needed in order to correctly reproduce the observations, including e.g., combination of optically thin and thick 
clouds, similarly as done in the visible range \citep{Binette96} or ionization by fast-shocks \citep{Dopita96}.

\subsection{The Origin of the H$_{2}$ Emission}

	Several molecular hydrogen emission lines are present in the K band spectra of our Seyfert 
galaxies.
The H$_{2}$~1-0S(1)~2.12~\micron\ is the strongest molecular transition in our spectra, 
followed by H$_{2}$~1-0S(2)~2.03~\micron.  
The H$_{2}$~1-0S(3)~1.96~\micron\ line is detected in all the galaxies, although in the case of NGC~7212 
is strongly blended with the [Si VI] coronal line. 
Both H$_{2}$~1-0S(0)~2.22~\micron\ and H$_{2}$~2-1S(1)~2.25~\micron~are marginally detected in practically all the sources, except 
in Mrk~1066 (see bottom panel of Figure \ref{fi:m1066}), the galaxy with the strongest stellar emission.

\begin{figure}[!ht]
\centering
\includegraphics[width=10cm,angle=90]{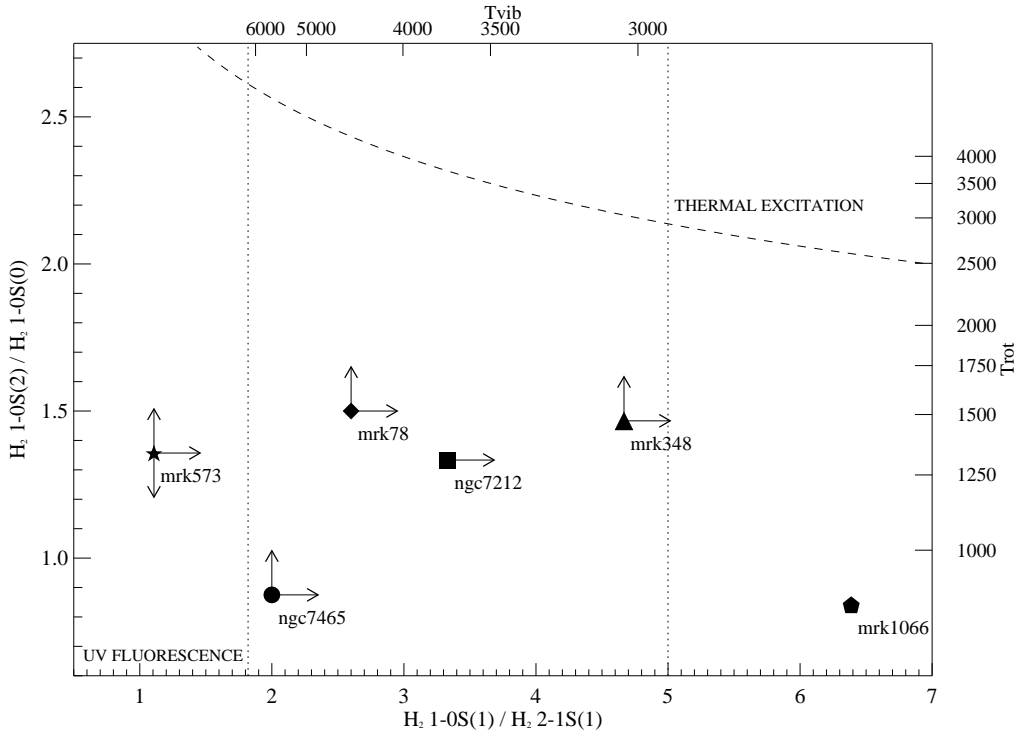}
\figcaption{\footnotesize{H$_{2}$ 1-0S(2)/1-0S(0) versus H$_{2}$ 1-0S(1)/2-1S(1) line ratios of the five galaxies in
our sample, together with nuclear Mrk~78 values \citep{Ramos06}. The dashed line indicates the locus
of equal  T$_{vib}$ and T$_{rot}$. In the case of Mrk~573, T$_{rot}$ is unconstrained, because
both H$_{2}$ 1-0S(2) and H$_{2}$ 1-0S(0) lines are upper limits. Vertical dotted lines represent the
regions of "thermal" and "non-thermal" excitation from \citep{Mouri94}. }
\label{hydrogen}}
\end{figure}

Basically there are two  H$_{2}$ excitation mechanisms: "thermal" and "non-thermal" processes or 
radiative decay from excited levels. In the thermal case the molecules are heated by shocks, UV photons
or X--rays, whereas in the non-thermal case the molecules are excited after absorption of an UV photon
or through collision with a fast electron from an X--ray ionized plasma.  
These two mechanisms produce different spectral features and the relative emission line intensities 
can be used to identify the dominant mechanism. In particular,
the H$_{2}$ 1-0S(1)/2-1S(1) line ratio is higher for thermal excitation (5-10) than for UV fluorescence
($\sim$1.82), as proposed by \citet{Mouri94}. 
Nevertheless, in the case of a dense gas (N$_{e}$ $\ge$ 10$^{4}~cm^{-3}$), collisional de-excitation 
of the H$_{2}$ molecule modifies the spectrum, approaching the thermal one.
The values obtained for the nuclear region of our galaxies  
tend to be closer to the case of thermal excitation (see Figure \ref{hydrogen}).
Excitation by hard X-rays is ruled out by other authors based on considerations of energetics 
\citep{Rodriguez04} and on the detection of the H$_{2}$ 1-0S(3)
transition in several Seyfert galaxies \citep{Davies05}, which would otherwise be suppressed in a
X-ray-irradiated gas.

Another way to discriminate between the thermal and fluorescent excitations is through the 
rotational and vibrational temperatures. We have determined both temperatures 
using the expressions given by \citet{Reunanen02}, using the H$_{2}$ 1-0S(1)/2-1S(1) and 
H$_{2}$ 1-0S(2)/1-0S(0) line ratios (see Figure \ref{hydrogen}). 
In the case of thermal excitation, both temperatures 
should be similar, whereas in the case of fluorescent excitation a high vibrational temperature 
must be in contrast with a lower rotational temperature. According to this criterium
the excitation of the H$_2$ molecule in the nucleus of our galaxies would be due to UV fluorescence, although
the line ratios H$_{2}$ 1-0S(1)/2-1S(1) are in most cases lower limits. 
The values of T$_{vib}$ for most of our galaxies are larger than those found by \citet{Reunanen02} for a
sample of Seyfert galaxies (between 1800 and 2700 K) and among the highest values found 
by \citet{Rodriguez04} for a similar group of AGN. 

The faintness of the lines involved in these diagnostics
introduces uncertainties that do not allow a more precise determination of the excitation mechanism. 
We tentatively conclude that thermal excitation is the most likely excitation mechanism in our galaxies, although 
UV-fluorescence surely plays its role. 

\section{Conclusions}

We presented and analyzed nuclear near-infrared spectra covering the range from 0.8 to 2.4 \micron~for 
five Seyfert galaxies, taken with the LIRIS/WHT facility. The main results are summarized in the following:

\begin{itemize}

\item The continua of the five galaxies considered in this work have been modelled with a combination of stellar templates 
plus a Blackbody component associated with hot dust emission (T=1000 K). This confirms that for the objects 
analyzed here the UV/optical power-law contribution to the nuclear 
continuum in the near-infrared range is negligible, contrary to what happens in Type-1 AGN.

\item The H and K ranges of the spectra are plagued of absorption features, that in some cases are highly diluted by 
non-stellar emission.  The stellar population of the five galaxies is dominated by late-type giants, with the average age
of the stellar systems being constrained between 100 Myr and 1 Gyr.

\item Different values of the optical extinction (A$_{V}$) among the galaxies have been derived from emission line ratios. 
Mrk~348 presents the most extinguished NLR (A$_{V}\sim$ 6 mag), whereas Mrk~573 and NGC~7465 have the lowest amount of extinction. 
It is noticeable that the two galaxies with detection of broad components in their emission line profiles present the less extinguished
NLRs.

\item High-ionization coronal lines with ionization potentials larger than 100 eV as [Si VI]$\lambda$1.963,
[Si X]$\lambda$1.430, and [S XI]$\lambda$1.920 \micron~are detected in Mrk~348, Mrk~573, and NGC~7212. On the contrary, 
for NGC~7465 and Mrk~1066 these transitions are extremely faint or undetectable, indicating their low-luminosity AGN nature.

\item Diagnostic diagrams involving iron and phosphorus forbidden transitions from the spectra of the galaxies considered allow us
to study the excitation mechanisms of the gas, indicating that not only photoionization, but also radio-jet interaction 
is taking place in the nuclear region of the galaxies. 

\item Several molecular transitions are detected in the K range of the spectra. The dominant 
excitation mechanism seems to be thermal excitation, although UV-fluorescence may be also present in 
some cases.

\item The detection of broad wings in the Pa$\alpha$ profile of Mrk~573 (FWHM$\sim$1200 km~s$^{-1}$) confirms the classification 
of this galaxy as an obscured NLSy1 galaxy, firstly claimed by \citet{Ramos08}.

\item The near-infrared spectrum of NGC~7465 confirms its LINER nature, lacking completely of high-ionization
lines, and presenting strong H$_{2}$ emission. 
The detection of broad wings in the profiles of Pa$\beta$ and Br$\gamma$ (FWHM$>$2000 km~s$^{-1}$) and
of Fe II and O I permitted lines allows to reclassify this galaxy as a  Type-1 LINER.

\end{itemize}

\appendix
\section{Individual objects}
\label{sc:indiv}

\subsection{Mrk~348 (NGC~262)}
\label{sc:m348}

The nucleus of Mrk~348 is optically classified as a Type-2 Seyfert. 
Hubble Space Telescope (HST) observations of this galaxy \citep{Capetti96,Falcke98}, show that 
the [O III] line emission
is confined to a linear structure of 0.45\arcsec~in size, oriented at a P.A. of 155$^{o}$. 
The good agreement between the radio and optical
structures \citep{Capetti96} supports the hypothesis of a nuclear engine located near the optical nucleus
and coinciding then with the apex of an ionization cone. 

From the X-ray point of view Mrk 348 shows an absorbed hard X-ray spectra 
with $\Gamma$=1.61$\pm$0.02, and the EW of its K$\alpha$ iron line similar to those of Seyfert 1 galaxies \citep{Awaki06},
plus a reflection component at higher energies \citep{Smith01}.
This feature classifies the galaxy as a Compton-thin Seyfert 2.  
X-ray variability in the continuum flux was found by \citet{Smith01}, while the iron K$\alpha$ line and
the reflection component did not vary. This suggest an scenario where the central X-ray source would be
surrounded by a clumpy distribution of absorbing material.

\begin{figure}[!ht]
\centering
\includegraphics[width=15cm]{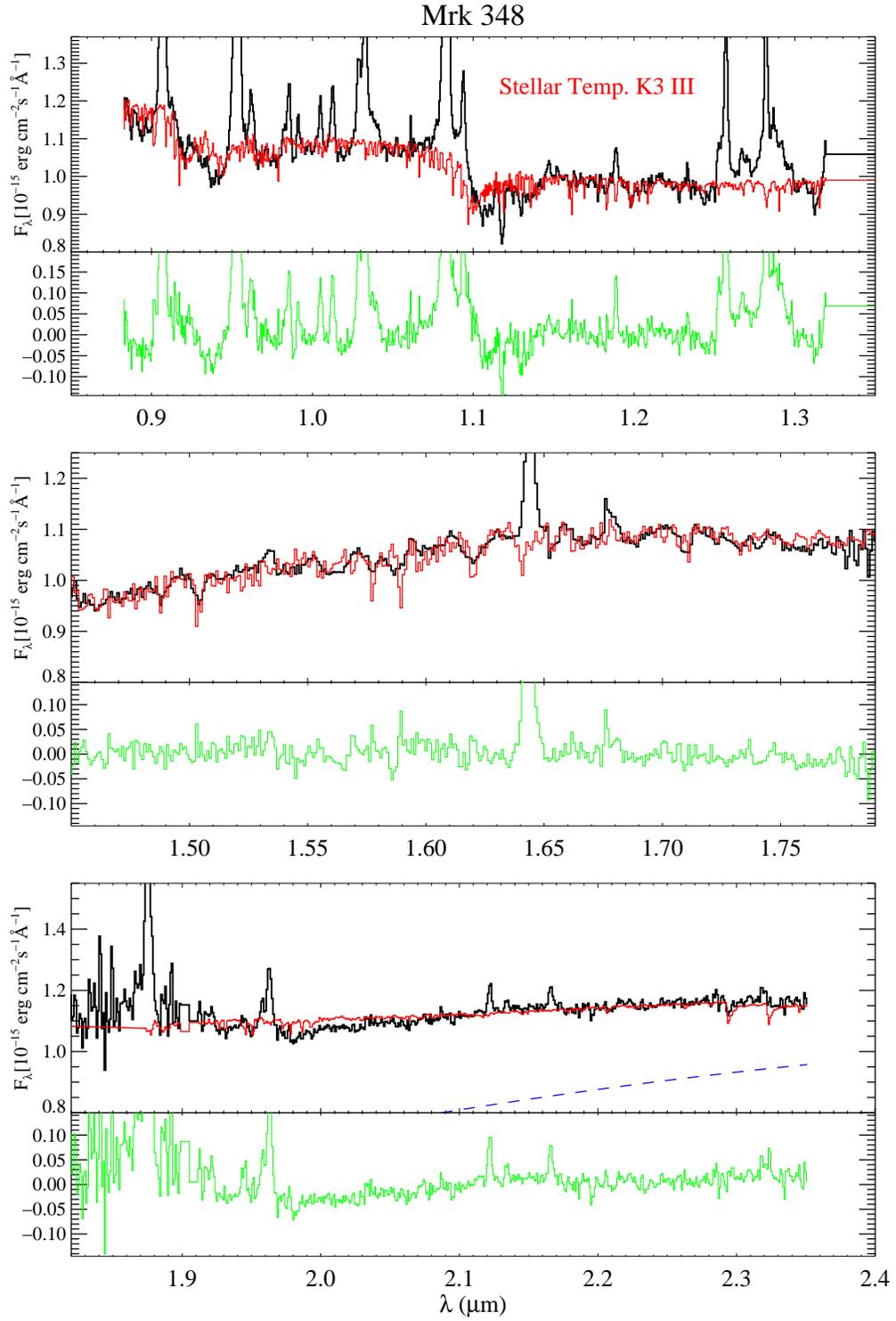}
\figcaption{\footnotesize{Nuclear spectrum of the galaxy Mrk~348 in the ZJ, H, and K ranges (in black) fitted with 
a stellar template (in red) and a BB component (in blue). The residuals resulting from the fit are also represented 
at the bottom of each panel.}
\label{fit348}}
\end{figure}

From our near-infrared spectrum, this galaxy appears as an archetypal Seyfert 2, showing prominent low- and 
high-ionization emission lines, together with a continuum flattening or even rising in the H and K
bands (see Figure \ref{sed}). The shape of its near-infrared continuum constitutes one of the 
most peculiar cases in our sample. 
It can be fairly reproduced by the combination of a stellar template of 
type K3~III plus a BB component (T$\simeq 1000$~K), which dilutes the stellar contribution by 
40\% and 77\% at the H and K bands, respectively (see Figure \ref{fit348}). 
In detail, there are few mismatches between 
the observed spectrum and the fitted model with respect to the stellar absorption features: 
the CO 2.29 \micron~band appears very diluted, whereas it is well marked in the model. 
The same happens with the CN feature at 1.1~\micron. The stellar template has to be 
reddened using a color excess E(B-V)=0.12. 
According to the EW ratio of CO~1.62~\micron\ to SiI~1.59~\micron~(see Table \ref{absorption}) 
the dominant stellar population in the H band of Mrk~348 corresponds to spectral types between K3 III 
and M3 III, which is consistent with the best fitting stellar template. 
On the other hand,  \citet{Gonzalez01} found that the stellar population present in Mrk~348 
is composed of old stars, with the absorption bands significatively diluted by the non-stellar continuum. 
They conclude that
the stellar population of the nuclei of this galaxy is similar to that of an elliptical galaxy,
but also with larger contribution of intermediate age stars.

The emission line spectrum presents few coronal or high ionization lines. The extinction
derived from the ratio of [Fe II] lines is the largest in the sample. Assuming that 
the coronal lines are produced in the inner part of the NLR, the high extinction implies that either there is a
screen in front of the NLR or it is already inmersed in dust. This dust distribution could be in turn 
related with the excess observed in the K band continuum due to thermal BB emission.

A broad H$\alpha$ component of FWHM $\sim$ 8400 km~s$^{-1}$ has been detected using  
optical spectropolarimetry \citep{Miller90,Tran95b}. However, we do not detect any 
broad emission wings in any of the recombination lines.

We have compared some line ratios with the predictions of photoionization models, that
are represented in Figure \ref{cloudy1}. The [S III]/Pa$\beta$ ratio is 
subestimated by the models, in comparison with the observed value for Mrk~348. The value 
of the ionization parameter seems to be log $U~\sim$ 0 by looking at the 
diagnostic diagram involving the [S III] and He II lines, but it is difficult to 
confirm it based in the [Fe II]/Pa$\beta$ versus  He I/Pa$\beta$ diagram.

According to the H$_{2}$ line ratios displayed in Figure \ref{hydrogen} for Mrk~348, the galaxy is 
located closer to the thermal excitation region. However, due to the uncertainties in the T$_{rot}$ 
and T$_{vib}$  calculations, that are lower and upper limits, respectively, 
UV fluorescence cannot be discarded.
            
Summarizing, the continuum of this Seyfert 2 galaxy is well reproduced by a stellar spectrum of type K3 III plus 
a BB of T$\sim$1000 K, showing a notable excess in the K band. The stellar absorption features present in the 
H and K bands are strongly diluted in comparison with the rest of the sample. According to the emission lines, 
Mrk~348 is the archetypal Seyfert 2, presenting prominent low- and high-ionization transitions. The nucleus of 
this source is the most extinguished of the sample with an A$_{V} \sim 6$ mag.

\subsection{Mrk~573}

This galaxy is known to have an extended emission-line region \citep{Unger87,Haniff88,Tsvetanov92}.
[O III] images from the HST reveal two ionization cones \citep{Pogge95,Capetti96,Falcke98}. 
The radio extended emission axis is aligned with the major axis of the NLR and misaligned 
with the the host galaxy major axis.
The high intensity of the [Fe II] lines is also indicative of the influence of the radio emission in the
gas excitation. 
In the X-ray regime this galaxy is a Compton-thick object that presents a soft X-ray excess 
and a steep photon spectral index 
\citep{Guainazzi05}, together with X-ray variability of a factor $\sim$2 within 300 s \citep{Ramos08}.

\begin{figure}[!ht]
\centering
\includegraphics[width=15cm]{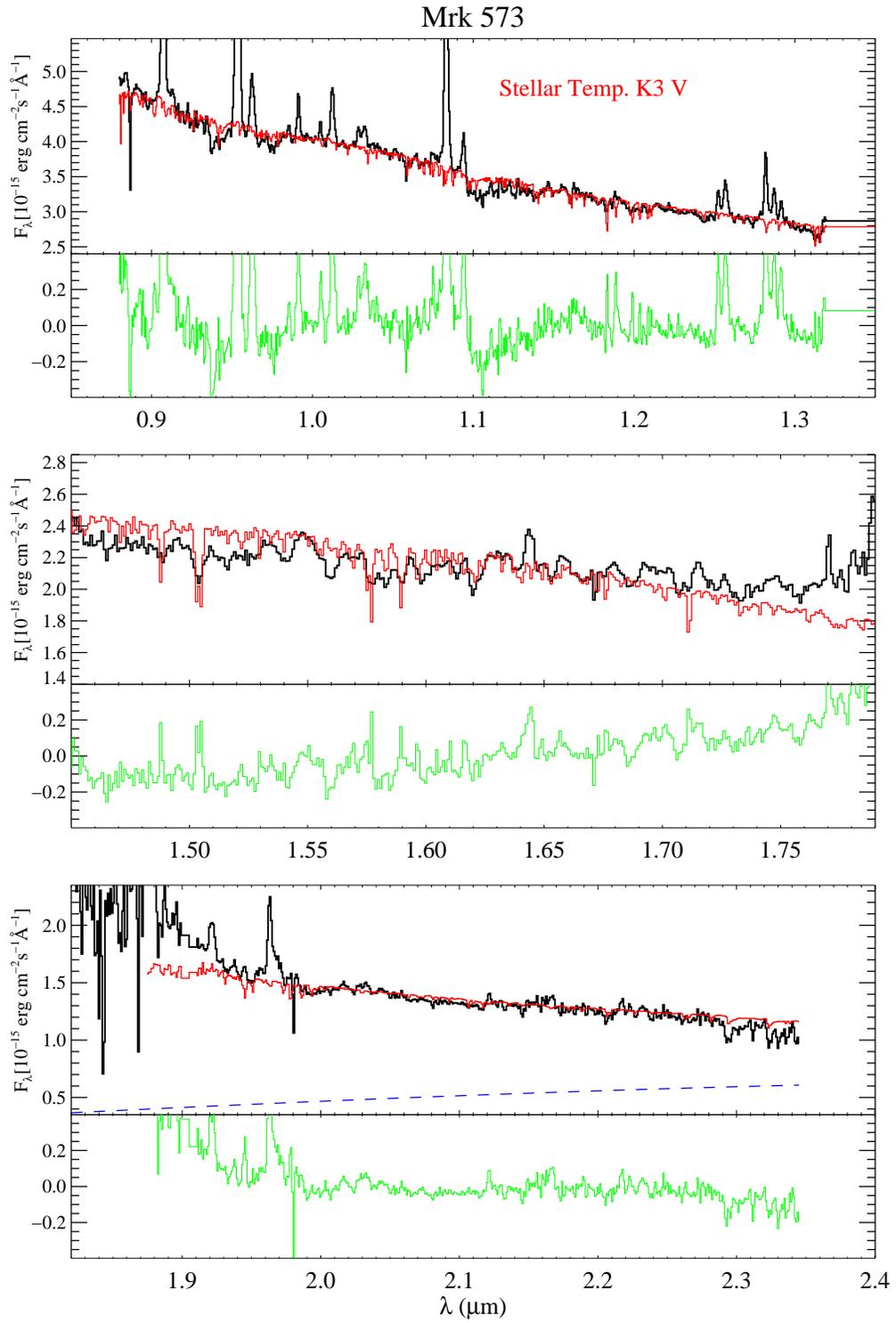}
\figcaption{\footnotesize{Same as in Figure \ref{fit348}, but for Mrk~573.}
\label{fit573}}
\end{figure}

The nuclear near-infrared spectrum of Mrk~573  presents very strong emission lines
together with prominent absorption features in the H and K bands.  
The shape of its continuum appears very steep and it can be well reproduced by a non-reddened stellar 
spectrum of type K3V (see Figure \ref{sed}) plus a BB contribution which dilutes the stellar 
spectrum by 15\% and 45\% at H and K bands, respectively (see Figure \ref{fit573}). 
The depth of some stellar features (the CN~1.1~\micron\ and the 
CO 2.29 \micron~bands) cannot be reproduced only by the dwarf+BB template, requiring the contribution of a giant or supergiant 
population.  Indeed, from the high CO 1.62 \micron~to Si I 1.59 \micron~ratio (see Table \ref{absorption}) the
derived population  corresponds to late-type supergiants K4 I and M1 I. \citet{Gonzalez01} found old stars 
in the nuclear region of Mrk~573, based on the EWs of the Ca II$\lambda\lambda$0.393,0.430 \micron~bands. 
As in the case of Mrk~348, they claimed
for a stellar population typical of an elliptical galaxy, with contribution of intermediate age stars. 
\citet{Rainmann03} determined that old population ($\sim$10 Gyr) is dominating the central parsecs of Mrk~573.

Mrk~573 has been classified by many authors as a Seyfert 2 galaxy, based on optical 
\citep{Tsvetanov92,Erkens97,Mullaney08} and infrared spectroscopy \citep{Veilleux97,Riffel06}. 
However, based on recent near-infrared spectroscopy in the ZJ range of Mrk~573, together with the detection 
of X-ray variability in the X-rays, \citet{Ramos08} have recently reclassified this object as an obscured NLSy1. 
The ZJ range of the spectrum (also reported here in Figure \ref{fi:m573}) shows permitted O I and Fe II lines 
well above the noise level. In addition, broad wings are detected in the Pa$\beta$ profile.
All these features can only be produced in a high-density optically thick gas, more internal than the NLR, and 
consequently, they are exclussive of Type-1 AGN and never seen in canonical Seyfert 2 galaxies.
In this work, we present also the HK spectral range of Mrk~573, that 
confirms the previously mentioned classification. Strong [Si X]$\lambda$1.430 and [Si VI]$\lambda$1.963 emission, 
together with the [S VIII]$\lambda$0.991, [S IX]$\lambda$1.252, and [S XI]$\lambda$1.920 lines reveal a high-ionization region, 
likely located closer to the nucleus than the classical NLR. 
The Pa$\alpha$ profile shows also a broad pedestal of 
FWHM$\sim$1200 km~s$^{-1}$. This component is narrower than the Pa$\beta$ broad component 
(FWHM$\sim$1700 km~s$^{-1}$), although both measurements are consistent within the errors (see Table \ref{fluxes}). 
Spectropolarimetric  observations \citep{Nagao04} reveal prominent scattered broad H$\alpha$ emission, 
together with the optical Fe II multiplet.
The measured [Fe II]1.257/1.644 \micron~line ratio for Mrk~573 indicates a negligible 
extinction towards the NLR, likely allowing to see more internal regions.

The comparison of the observed line ratios with the predictions from photoionization models does not show
a good matching with any of the inspected models by looking at the He II/Pa$\beta$ versus [S III]/Pa$\beta$ diagram.
It appears that either a harder ionizing continuum is needed or 
the line flux of the narrow component of $\mathrm{Pa}\beta$ is underestimated.
On the contrary, the HeI/Pa$\beta$ and [Fe II]/Pa$\beta$ ratios are nicely reproduced by the models, 
using a hydrogen density of n$_{H}$ = 10$^{5}$ cm$^{-3}$ and a log $U~\in$ [-1.5,-1].  

With respect to the molecular content, the H$_{2}$ emission is very weak in Mrk~573, 
and the most likely excitation mechanism is UV fluorescence according to the H$_{2}$ 1-0S(1)/2-1S(1) ratio (see Figure \ref{hydrogen}). 

In conclusion, the steep continuum of this obscured NLSy1 galaxy can be reproduced with a stellar spectrum of 
type K3 V with the extinction of the stellar spectrum being negligible. The dilution in the H and K bands is intermediate,
compared with the rest of the sample, and the emission lines are the most prominent. Strong coronal lines are detected 
in the spectrum, revealing a high-ionization region, likely closer to the nucleus than the NLR. Simple photoionization
models are unable to reproduce accurately the line ratios. From the intensity of [Fe II] lines it seems very likely that
the interaction with radio emission plays a role.

\subsection{Mrk~1066}
\label{sc:m1066}

\citet{Bower95} reported low-excitation emission line ratios similar to those of LINERs for this Seyfert 2 galaxy, 
from ground-based spectroscopy. 
Spectropolarimetric observations of Mrk~1066 have not revealed broad components in the permitted lines 
\citep{Miller90}. The results of the observations are consistent with the polarization being almost 
completely due to foreground sources. 
\citet{Ulvestad89} mapped the linear nuclear radio source finding a jetlike morphology elongated with a P.A.
= 134$^{o}$, close to the [O III] axis.  The position angle of the optical ionization cones is
135$^{o}$. 
Hubble Space Telescope (HST) Wide Field Planetary Camera 2 (WFPC2) and VLA radio imaging were employed by
\citet{Bower95} to analyze the distributions of the emission-line gas and radio continuum, that are consistent
with a bipolar jet significantly inclined with respect to the galaxy disk. 
From an X-ray perspective, this galaxy has been classified by 
\citet{Shu07} as a 
Compton-thick object on the basis of its large Fe K$\alpha$ EW ($>$ 1 keV). 
The power-law photon index reported by these authors is 2.75$\pm^{0.17}_{0.07}$.
On the other hand, it is a very luminous object in the far-infrared, 
showing a double nucleus \citep{Gimeno04}.

\begin{figure}[!ht]
\centering
\includegraphics[width=15cm]{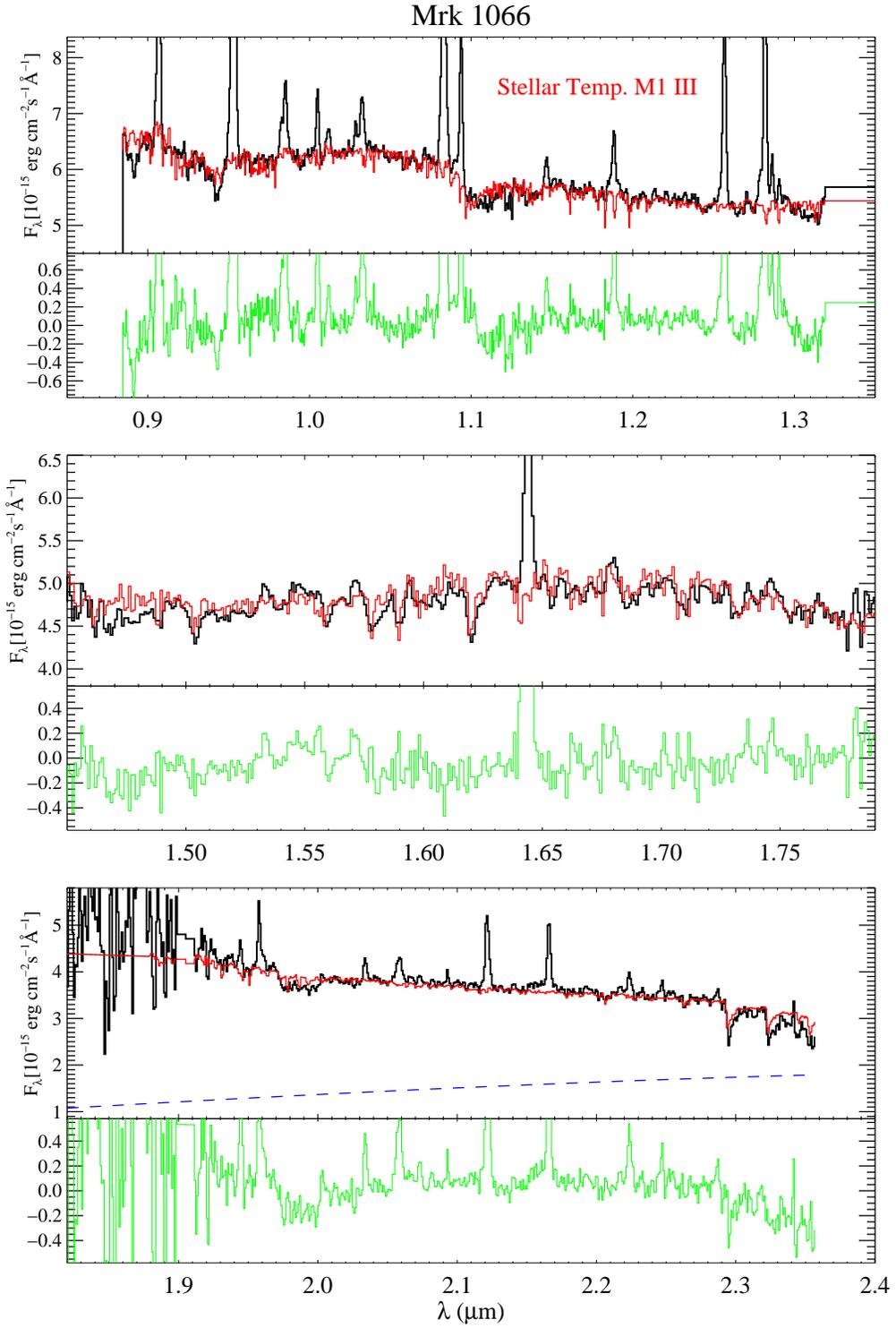}
\figcaption{\footnotesize{Same as in Figure \ref{fit348}, but for Mrk~1066.}
\label{fit1066}}
\end{figure}

From our near-infrared spectrum, Mrk~1066 seems to be a Seyfert 2 with relatively weak AGN 
signatures. The High-ionization lines are faint in comparison with the recombination
lines.  The strength of the  Br$\gamma$ and H$_{2}$ transitions are indicative of 
vigorous star formation activity. As an example, we note that 
the H$_{2}$1-0S(3) line is $\sim$4 times more intense than the [Si VI]$\lambda$1.963 feature. 
Its nuclear continuum is the steepest in our sample, showing the deepest 
stellar features. 
According to its CO 1.62 \micron~to Si I 1.59 \micron~ratio (see Table \ref{absorption}), the 
stellar population responsible of the absorption bands is dominated by K3 III stars. The dilution
fraction derived from the absorption features is negligible in the H band. 
\citet{Gonzalez01} found that the stellar population in Mrk~1066 is compatible with young and 
intermediate age stellar systems. 
\citet{Rainmann03} claim for a 1 Gyr stellar population dominating the nuclear region of this galaxy.
On the other hand, the shape of the continuum of Mrk~1066 can be well reproduced by a stellar 
template of type M1~III (see Figure \ref{fit1066}). The  depth of the CO 2.29 \micron\ feature in the M1~III template 
is  not as deep as observed, indicating the presence of supergiant stars or possible 
starburst activity. The CN 1 \micron\ feature appears very strong and it is reproduced very approximately 
by the stellar template (see Figure \ref{fit1066}). 
From the best fitting model (stellar template plus BB) the required extinction of 
the stellar spectrum is E(B-V)=0.12 (compatible with Galactic extinction) and the dilution fractions are
14\% and 48\% in the H and K bands, respectively.

The extinction derived from the [Fe II]1.257/1.644~\micron\ line ratio (see Figure \ref{reddening}) 
is among the highest ($A_V = 4$ mag) in our sample. 
The observed line ratios shown in Figure \ref{cloudy1} are not perfectly reproduced 
with the photoionization models obtained with \textsc{Cloudy} for the case of Mrk~1066.

Mrk~1066 is the object in our small sample of galaxies with more intense H$_{2}$ emission. As it can be
seen from Figure \ref{hydrogen}, the H$_{2}$ 1-0S(1)/2-1S(1) line ratio locates the galaxy in the thermal
excitation region. Nevertheless, the difference between its rotational temperature (T$_{rot}$=850$\pm$170 K)
and the vibrational temperature (T$_{vib}$=2600$\pm$400 K) seems to indicate that 
thermal excitation is not the only exciting mechanism.

Summarizing, the shape of the continuum (the steepest of the sample) is reproduced 
by a stellar spectrum of type M1 III.
Nevertheless, the CO 2.23 \micron~feature is deeper than the observed for this type of stars, indicating the presence of 
supergiant stars in the nuclear region of the galaxy, where likely starburst activity is taken place. A moderate amount 
of dilution is affecting the H and K absorption features. 
High-ionization lines are very weak in comparison with the recombination lines. The results 
obtained from the comparison with photoionization models indicate low values of the ionization parameter. Both facts 
point towards a weak-luminosity AGN.
The NLR of Mrk~1066 presents a considerable amount of extinction.
The observed H$_{2}$ emission is the most intense of the sample, and the more likely excitation mechanism is thermal excitation.

\subsection{NGC~7212}
\label{sc:n7212}

NGC~7212 is part of a triple interacting system \citep{Wasilewski81}.
HST [O III] images reveal that the emission is extended along a P.A. = 170$^{o}$ and it is composed 
of several individual blobs towards the north and south of the nucleus \citep{Schmitt03}. 
This [O III] emission is aligned with the radio emission of the galaxy. 
Several dust lanes are revealed in the continuum images \citep{Falcke98}, together with
the elongated and diffuse NLR structure, with a strong central peak near the continuum maximum. 
The VLA maps show a double compact radio source orientated in the North-South direction. 
Spectropolarimetric observations reveal a broad H$\alpha$ component of FWHM $\sim$ 4000 km~s$^{-1}$ 
\citep{Tran92,Tran95a}.
At the X-ray range, its spectrum (Chandra/ACIS) have been fit by a pure Compton reflection component
and a strong K$\alpha$ iron line, suggesting that the source is Compton-thick \citep{Bianchi06}. 

\begin{figure}[!ht]
\centering
\includegraphics[width=15cm]{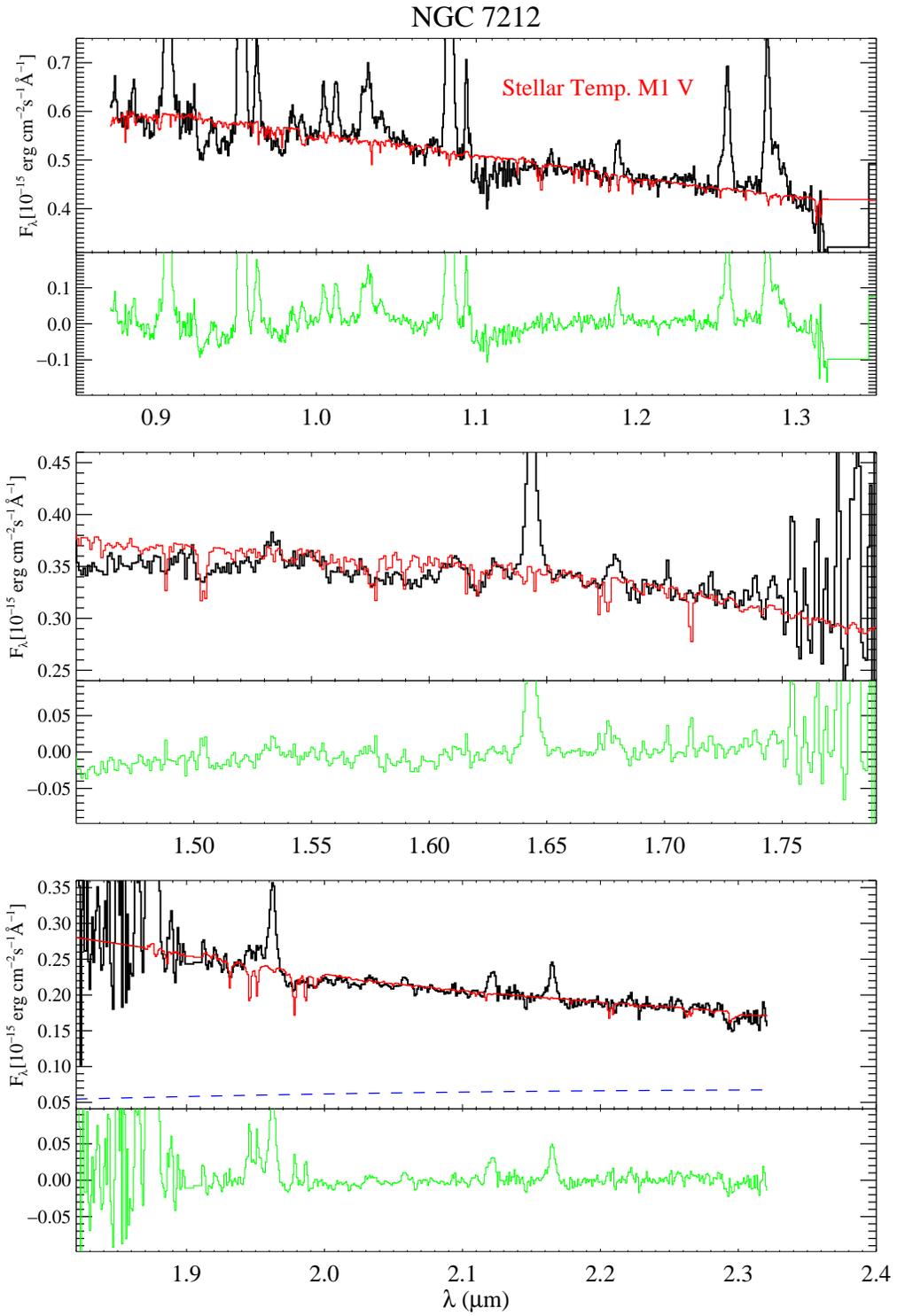}
\figcaption{\footnotesize{Same as in Figure \ref{fit348}, but for NGC~7212.}
\label{fit7212}}
\end{figure}

From our near-infrared data, NGC~7212 presents the typical spectrum of a Seyfert 2 galaxy, similar to 
that of Mrk~348, but without the K band excess. Its continuum is very flat (see Figure \ref{sed}) 
and several narrow emission lines are present in the spectrum, including all the high-ionization 
species seen in Mrk~573. The [S IX]$\lambda$1.252 line is strongly blended with the 
[Fe II]$\lambda$1.257 feature in NGC~7212. The best fit of the continuum was found with 
a stellar template of type M1V plus a relatively small BB contribution in the 
K band (see Figure \ref{fit7212}). The required extinction of the stellar template is negligible and the dilution 
fractions are 0\%  and 20\% at H and K bands, respectively. Nevertheless, the depth of the CN 1 \micron\
feature is better reproduced by a M1~III template.
\citet{Gonzalez01} and \citet{Rainmann03} found the same old stellar population as in Mrk~348 
and Mrk~573, described above.

The amount of extinction derived from the [Fe II]1.257/1.644~\micron\ line ratio is about 
$\mathrm{A}_{V} \sim 2$. 
The line ratios considered in Figure \ref{cloudy1} for their comparison with 
photoionization models are nicely reproduced for the case of NGC~7212, using an ionization parameter 
log $U$ $\sim$ 0. Thus, photoionization seems to be confirmed as the dominant mechanism in the nucleus of this galaxy.

The vibrational and rotational temperatures determined for the case of NGC~7212 are 
considerably different, with $\mathrm{T}_{vib} >> \mathrm{T}_{rot}$, which is typical of UV fluorescence. 
Nevertheless, its H$_{2}$1-0S(1)/2-1S(1) ratio is in
the intermediate region between the thermal excitation and UV pumping values, making difficult 
to determine which is the H$_{2}$ excitation mechanism dominating in this galaxy.

In summary, the flat continuum of NGC~7212 is reproduced by a stellar spectrum of type M1 V, although the 
depth of the CN 1.1 \micron~feature is better fitted using a M1 III template. The extinction of the stellar spectrum is negligible, and
the dilution factor in the H and K bands in the lowest in the sample. The emission line spectrum corresponds to a typical Seyfert 2
galaxy, very similar to that of Mrk~348, but without the red excess. Several narrow emission lines populate the spectrum, including 
high-ionization species. The comparison with photoionization models indicate that the considered line ratios can be 
reproduced by this type of ionization mechanism.

\subsection{NGC~7465 (Mrk~313)}
\label{sc:n7465}

\citet{Osterbrock87} and \citet{vanDriel92} claimed for the low-ionization AGN nature 
of the nucleus of NGC~7465, based on optical spectroscopy. Its spectrum fullfils the criterium of having 
[O II]$\lambda$3727~\AA\ at least as strong as [O III]$\lambda$5007~\AA\ 
\citep{Heckman80}\footnote{\citet{Osterbrock87} reported [O II]$\lambda$3727 / [O III]$\lambda$5007 = 1.4.},
but the ratio [O I]$\lambda$6300 / [O III]$\lambda$5007 = 0.1 \citep{Osterbrock87} appears 
too small for a classical LINER. 
Nevertheless, \citet{Filippenko92} defined the weak-[O I] LINERs as those having 
$\mathrm{[O~I]}\lambda 6300 / \mathrm{H}\alpha < 0.17$.
The reported value of this ratio for Mrk~7465 by \citet{Osterbrock87} is 0.1, which
classifies it in this LINER subclass.
The hard X-ray absorption-corrected luminosity of NGC~7465 is 10$^{42}~erg~s^{-1}$, 
according to \citet{Guainazzi05}, that is within the typical LINER intrinsic 
luminosities in the 2-10 keV band (10$^{40-42}$ erg~s$^{-1}$; 
\citealt{Terashima00}).

Our near-infrared spectrum confirms the LINER nature of this galaxy, that lacks completely of the high 
ionization lines [S IX]$\lambda$1.252, [Si X]$\lambda$1.430, and [S XI]$\lambda$1.920.
[S VIII]$\lambda$0.991 and [Si VI]$\lambda$1.963, that are usually very prominent 
in Seyfert galaxies, are only marginally detected in NGC~7465.

\begin{figure}[!ht]
\centering
\includegraphics[width=15cm]{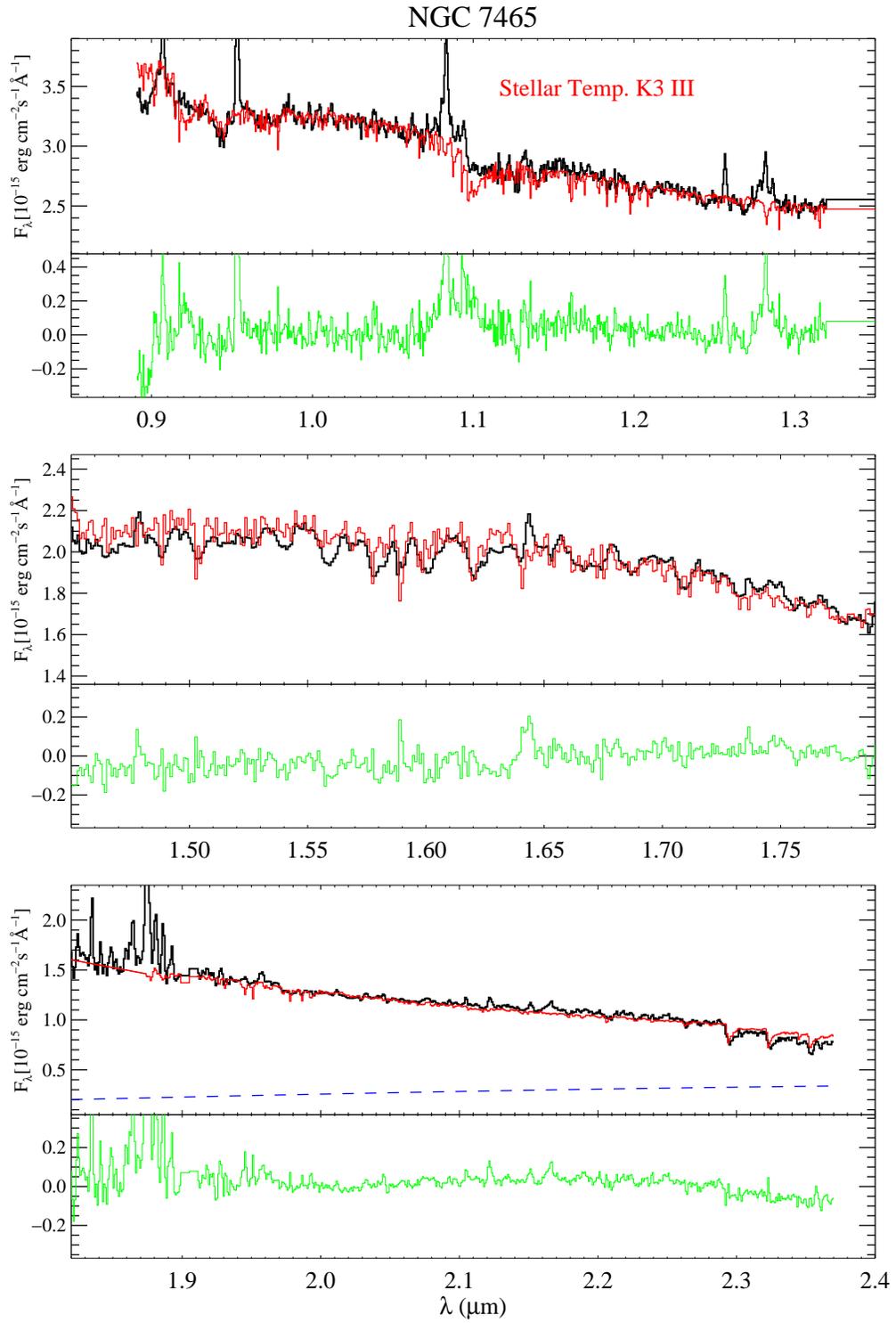}
\figcaption{\footnotesize{Same as in Figure \ref{fit348}, but for NGC~7465.}
\label{fit7465}}
\end{figure}

The dominant stellar population in the nuclear region of NGC~7465 corresponds to types between K3 III and M3 III, 
according to relative absorption bands measurements (see Table \ref{absorption}). 
The spectral shape in the near infrared range is relatively steep, similar to that of Mrk~573 (see Figure \ref{sed}),
and it can be well reproduced by a stellar template of type
K3 III (see Figure \ref{fit7465}). The depth of the CN 1 \micron~and CO~2.29~\micron\  features are well matched between the template
and the observations. The dilution fractions due to the BB component are 6\% and 32\% at H and K
bands, respectively.

Apart from confirming NGC~7465 as a low-ionization AGN, our near-infrared emission line spectrum 
reveals for first time the Type-1 AGN nature of this galaxy, based on the detection of  broad
components of Pa$\beta$ and Br$\gamma$ (with FWHM $>$ 2000 km~s$^{-1}$), together with Fe II and O I permitted 
lines (see Figure \ref{fi:n7465}). The broad component of Pa$\alpha$ cannot be properly measured because of its 
coincidence with a strong telluric absorption band.
The width of the lines in the optical has been measured with different results: 
\citet{Osterbrock87} found very narrow H$\alpha$ and H$\beta$ profiles ($\sim$250 km~s$^{-1}$), 
while \citep{Maehara88} reported on emission lines as broad as 600 km~s$^{-1}$.
The detection of broad wings in the recombination lines of LINERs
(reported for first time by \citealt{Heckman80}), indicates that the photoionizing 
source is an AGN, and not a strong stellar emission, as it seems to happen in a different type of LINERs \citep{Alonso00}.
Moreover, the presence of O I and Fe II lines\footnote{These lines are 
produced by Ly$\beta$ and Ly$\alpha$ pumping in a fluorescence mechanism, respectively.}
suggests that strong UV emission is present, as well as high-density clouds similar to those of the BLR. 

The nuclear region of NGC~7465 is little extincted ($\mathrm{A}_{V}=1.1$ mag) according to its 
[Fe II]1.257/1.644 \micron~line ratio (see Figure \ref{reddening}). In addition, the nuclear
spectrum of NGC~7465 shows the highest value of [Fe II]$\lambda$1.257/Pa$\beta$. 
\citet{Larkin98} proposed two different
classes of LINERs according to this ratio: weak-[Fe II] LINERs (those with [Fe II]$\lambda$1.257/Pa$\beta$ $<$ 2) and 
strong-[Fe II] LINERs ([Fe II]$\lambda$1.257/Pa$\beta$ $>$ 2). The former class would be low-luminosity AGN,
whereas the latter would be powered by starburst. According to this classification, NGC~7465 would be a 
weak-[Fe II] LINER, consistent with being a low-luminosity AGN as derived from its X-ray luminosity.

The line ratios considered in Figure \ref{cloudy1} for the comparison with photoionized models computed 
with \textsc{Cloudy} are perfectly reproduced using values of the ionization parameter log $U~\sim$ -0.5 and 
a hydrogen density of 10$^{3}$ cm$^{-3}$. This indicates that photoionization alone permits to 
explain the measured values of the considered lines. 

In comparison with the rest of our sample, NGC~7465 has the strongest H$_{2}$ emission 
(see Table \ref{fluxes}), that is also typical of LINERs \citep{Larkin98}. 
Its H$_{2}$ 1-0S(1)/2-1S(1) line ratio is near the typical values for UV fluorescence (see Figure
\ref{hydrogen}). Concretely, this case is similar to that of Mrk~504 \citep{Rodriguez04}.
The T$_{vib}$ and T$_{rot}$ differ appreciably, indicating the presence of non-thermal
excitation. 
The low value of T$_{rot}$ can be attributed to UV heating and this process can be the main source of the 
H$_{2}$ excitation. 
UV heating models \citep{Black87}, where FUV photons emitted by OB stars
(photodissociation regions) heat the gas, predict excitation temperatures of 1000 K for the H$_{2}$ 
thermal component. Such low values are typically found in starburst galaxies \citep{Mouri94}.
Consequently, the molecular spectrum of NGC~7465 would be produced by UV fluorescence, according to its 
H$_{2}$ 2-1S(1)/1-0S(1) line ratio and vibrational and rotational temperatures.

Summarizing, the shape of the continuum of this Type-1 LINER is well reproduced by a stellar spectrum of type K3 III, with the 
depth of the CO 2.23 \micron~feature correctly matched. A moderate amount of dilution is afecting the absorption bands seen in the 
H and K bands of the galaxy. 
Hydrogen recombinations lines are also very prominent in the nucleus of this galaxy,
probably due to star formation. The comparison with photoionization models
confirms that the photoionizing source is a low-luminosity AGN, and not a strong stellar continuum, as it would
happen in other type of LINERs. In addition, the [Fe II]$\lambda$1.257/Pa$\beta$ ratio classifies the galaxy as a weak-[Fe II] LINER.

\clearpage

\acknowledgments

This work was partially funded by PN AYA2007-67965-C03-01, PN AYA2004-03136, and PN AYA2005-04149.

The William Herschel Telescope and its service programme are operated on the island of La Palma by the Isaac Newton Group in the
Spanish Observatorio del Roque de los Muchachos of the Instituto de Astrof\'\i sica de Canarias.

The authors acknowledge the data analysis facilities provided by the Starlink Project, which is run by CCLRC
on behalf of PPARC.

The authors acknowledge Elena Puga  Antol\'\i n and Mar\'\i a Rosa Zapatero Osorio 
for their valuable help during observations.

We finally appreciate the very useful comments of the anonymous referee.

\end{document}